\documentclass[acmsmall]{acmart}

\AtBeginDocument{%
  }

\setcopyright{acmlicensed}
\copyrightyear{2026}
\acmYear{2026}
\acmDOI{XXXXXXX.XXXXXXX}
\acmConference[Conference acronym 'XX]{Make sure to enter the correct
  conference title from your rights confirmation email}{June 03--05,
  2018}{Woodstock, NY}
\acmISBN{978-1-4503-XXXX-X/2018/06}

\begin{document}

\title{Remote Triggers: Misophonia, Technology Non-Use, and Design for Inclusive Digital Spaces}

\author{Tawfiq Ammari}
\email{tawfiq.ammari@rutgers.edu}
\orcid{0000-0002-1920-1625}
\affiliation{%
  \institution{Rutgers University}
  \city{New Brunswick}
  \state{NJ}
  \country{USA}
}

\author{Samantha Gilgan}
\email{larst@affiliation.org}
\affiliation{%
  \institution{Rutgers University}
  \city{New Brunswick}
  \state{NJ}
  \country{USA}
}

\renewcommand{\shortauthors}{Ammari and Gilgan}

\begin{abstract}
Misophonia, characterized by intense negative reactions to specific sounds or related visual cues, remains poorly recognized in clinical settings yet profoundly affects daily life. This study examines how individuals with misophonia experience and sometimes avoid technology that amplifies their triggers. Drawing on 16 semi-structured interviews with U.S. adults recruited from online communities, we explore how social media platforms such as TikTok and Instagram, along with remote communication tools like Zoom and Discord, shape coping strategies and patterns of non-use. Participants described frequent distress from uncontrollable audiovisual content and food-related behaviors during virtual gatherings. We propose design interventions—including channel-specific audio-visual controls, real-time trigger detection, and shared preference tools—to better support misophonic users and reduce exclusion in increasingly mediated social and professional contexts.
\end{abstract}



\received{20 February 2026}
\received[revised]{12 March 2026}
\received[accepted]{5 June 2026}

\maketitle

\section{Introduction}
\textit{Imagine attending a three-hour virtual team meeting where a colleague eats lunch on camera. For most participants, this might be barely noticeable or mildly distracting. But for someone with misophonia, the sound and sight of chewing triggers an involuntary fight-or-flight response—racing heart, sweating, overwhelming anger or panic (see examples here \cite{fulton_2019_misophonia}). The only options: endure physiological distress, mute everyone and miss the conversation, or leave the meeting entirely.}

Misophonia is a contested disorder with no clear definition or aetiology \cite{swedo2022consensus} that could be associated with a number of auditory disorders (e.g., hyperacusis) or neurodivergence (e.g., autism spectrum disorder) though not strictly associated with any of these conditions. Misophonia is characterized by abnormally strong reactions — such as anger and disgust — to a specific sound or pattern of sounds \cite{cavanna2015misophonia,potgieter2019misophonia}. It can also manifest in visual cues closely associated with the sources of the audio triggers \cite{jaswal2021misokinesia,webb2022beta}. It is a relatively newly
recognized condition, and therefore does not appear in the Diagnostic and Statistical Manual of
Mental Disorders \cite{schroder2013misophonia}. It has been in contention within the field of psychology whether misophonia
should be categorized as a mental disorder or not; arguments against it claim that it will further be
stigmatized and cause overdiagnosis of commonplace quirks as mental disorders, while others
argue that classification of misophonia as a mental disorder in the DSM will allow for raised public
awareness of it and validation to those who suffer from it \cite{taylor2017misophonia}. However, mental disorders are
characterized by the National Institute of Health as any disturbance in an individual's cognition,
emotion regulation, or behavior that causes dysfunction in a psychological, biological or
development process that can in turn affect mental functioning and causes a clinically significant
disturbance \cite{stein2021mental}. They often are associated with significant distress or diminished ability in social,
occupational, and other important situations. For individuals with misophonia (commonly referred to as misophones) hearing a trigger sound
creates an emotional or physical reaction, or both, that produces an inability to function in a
situation until (and often after) a trigger sound stops. Unlike a mild annoyance or dislike of certain
sounds, individuals with misophonia's fight-or-flight response is triggered upon hearing a trigger
sound and alters them unable to function in an environment, or may cause them to avoid said
situations all together. These reactions can completely alter individuals' social and professional
spheres.

Fricker defines epistemic injustice as the inability to effectively convey one's lived experience to those who lack a similar frame of reference \cite{fricker_epistemic_2007}. The necessity of repeatedly recounting these events can be re-traumatizing, resulting in epistemic trauma. As the text notes, "After so [many] repeated instances of silencing, she begins to doubt her own experiences... and eventually no longer feels comfortable sharing" \cite[P. 45-46]{stinnett2018trauma}. This harm is amplified by power asymmetries, particularly in interactions between patients and medical professionals. Mohammed notes that this form of trauma can impact both the individuals sharing their stories and the professionals receiving them, such as researchers \cite{mohamed2024debilitating}. Central to trauma-informed design \cite{chen_trauma-informed_2022, Scott2023} is the prioritization of user agency and voice. These frameworks provide a method for applying the Substance Abuse and Mental Health Services Administration’s (SAMHSA) principles to digital spaces. Specifically, they help examine how social media affordances—such as algorithmic recommendations and comment sorting \cite{parchoma_contested_2014}—and governance structures like moderation \cite{Jhaveretal2018} can be adapted to better support online communities.

Earlier work by Race et al. \cite{race_et_al_21} proposes auditory design recommendations for neurodivergent-aware technology and Das et al. \cite{das_et_al_21} study the challenges faced by neurodivergent individuals when working remotely, especially when using technologies like Zoom. The broader HCI literature on neurodivergent technology use reveals both affordances for connection and barriers to participation. DeVito et al. \cite{jiang2025shifting} document how neurodivergent individuals are often positioned as consumers rather than creators of content, noting that ``when social computing systems designed for neurodivergent people rely upon a conception of disability that restricts expression for the sake of preserving existing norms surrounding social experience, the result is often simplistic and restrictive systems that prevent users from `being social' in a way that feels natural and enjoyable.'' Jiang et al. \cite{jiang2025shifting} contribute insights on how people with ADHD experience video content, finding that ``participants faced both overstimulation and understimulation from visuals and audio (e.g., flashing lights, slower speech), which impacted their attention, engagement, and information retention.'' Sabinson \cite{sabinson2024pictorial} argues for ``supporting self-determined behaviors instead of aiming to `fix' behaviors deemed abnormal or problematic by society,'' proposing ``more fidgets and fewer fixes'' as a guiding principle.

Technology non-use is not monolithic; rather, it encompasses a spectrum of engagement patterns that reflect users' strategic negotiations with technology. Selective use represents ``partial domestication'' where users engage with specific affordances while rejecting others. Salovaara et al. describe this ``appropriation work'' \cite{balka2006making, salovaara2011everyday} where users repurpose technology (e.g., using smartphones but disabling social notifications) to prevent technology from becoming intrusive. This aligns with Sørensen's view that domestication involves users disciplining artifacts to serve the household's moral economy \cite{sorensen2006domestication, silverstone1992information}. For individuals with misophonia, such selective use might manifest as engaging with video platforms but muting audio, participating in text-based communication while avoiding voice calls, or using social media during specific low-trigger times.

While standard domestication implies a permanent state of use, episodic use involves cycles of starting and stopping. Rather than viewing these gaps as failures, scholars now frame them as a "self-care" mechanism \cite{gorm2019episodic}. When constant data monitoring becomes a "burden" \cite{epstein_2020} that causes anxiety, users "pause" their use of the technology. This results in a flexible, cyclical process where devices are adopted for specific purposes, set aside when they become overwhelming, and re-integrated whenever they are needed again. For misophones, this pattern is particularly salient: periods of platform engagement may be followed by extended withdrawal when trigger exposure becomes overwhelming, only to return when social or professional necessity demands re-engagement.

Building on these perspectives that center neurodivergent individuals' agency in design \cite{sabinson2025blowfish,lukava2022two}, we investigate how misophonia shapes technology use and non-use across social, professional, and leisure contexts. Unlike prior work that documents challenges, we focus on the design implications of these experiences, asking how platforms might better support sensory diversity. In this study, we ask:
\begin{quote}
    \textbf{RQ1: How does misophonia shape patterns of technology use and non-use across social media, video conferencing, and leisure platforms?}
    
    \textbf{RQ2: What coping strategies do individuals with misophonia employ to manage technology-mediated trigger exposure?}
    
    \textbf{RQ3: What design principles can guide the development of sensory-accessible platforms that accommodate misophonic users without requiring full withdrawal from digital participation?}
\end{quote}

We draw from 16 semi-structured, qualitative interviews with individuals with misophonia to expand Das et al.'s \cite{das_et_al_21} characterization of audiovisual challenges faced by neurodivergent workers to show how similar challenges can also appear in other life avenues. Our findings reveal that technology non-use among misophones spans from selective feature avoidance to complete platform abandonment, shaped by both auditory and visual triggers (misokinesia). We contribute a design framework with five principles—disaggregated audiovisual control, real-time sensory filtering, sensory predictability, collaborative boundary negotiation, and visual trigger management—that address the systemic exclusion of sensory-sensitive users from increasingly mediated social and professional life.

\section{Related Work}

We position our work at the intersection of three critical areas: sensory accessibility and neurodivergent design, participatory methods with disabled communities, and critical platform studies. Rather than surveying literature comprehensively, we build an argument for why current approaches fail sensory-sensitive users and how design research can address these gaps.

\subsection{Sensory Accessibility in Digital Environments}

Sensory sensitivities affect a substantial portion of the population, particularly among neurodivergent individuals \cite{poulsen2024auditory}. Recent work has begun examining how technology can support or hinder sensory wellbeing. Race et al. \cite{race_et_al_21} propose auditory design recommendations including attention to sound duration, wavelength, intensity, and recognizability. Zolyomi and Snyder \cite{zolyomi2021social} develop a sensory design map informed by neurodivergent experiences, highlighting the importance of customizable sensory environments.

However, this work primarily addresses designed sensory experiences (e.g., notification sounds, haptic feedback) rather than user-generated content and social behavior mediated by platforms. Das et al.'s \cite{das_et_al_21} examination of neurodivergent professionals' remote work practices comes closest to our focus, documenting challenges with ambient noise and lack of control over sensory input in video meetings. We extend this work from professional contexts to social and leisure platforms, and from problem documentation to design intervention.

Jiang et al. \cite{jiang2025shifting} contribute novel insights on how people with ADHD experience video content, finding that participants ``desired adjustable sound channels for aiding focus, video summaries for retaining information, and warnings for preempting sensory discomfort.'' Notably, participants found that ``redundant visual information, like captions, supported their auditory information processing'' \cite{jiang2025shifting}. Their work proposes three ADHD-inclusive design principles: ``(1) Adapt to ADHDers' levels of focus, (2) Utilize multisensory elements with flexibility to adjust to ADHDers' thresholds, and (3) Involve people with ADHD throughout the design process'' \cite{jiang2025shifting}. These principles challenge the misconception that ADHDers have ``a complete inability to focus,'' instead framing the experience as ``a mismatch between external expectations and differences in processing external input and information'' \cite{jiang2025shifting}. The ``curb cut effect'' suggests that such accommodations often benefit all users, pointing toward broader relevance of sensory accessibility features.

A critical gap remains: existing recommendations focus on what designers should avoid or include, but offer limited guidance on supporting \textit{diverse sensory needs simultaneously}. When one person's engaging ASMR content is another's distressing trigger, design must move beyond one-size-fits-all solutions to individualized, user-controlled sensory management.

Recent work on audiovisual accessibility for people with complex communication needs (CCNs) offers promising directions. Nevsky et al. \cite{nevskyetal25} explore ``highly personalised'' accessibility interventions for people with aphasia, arguing that audiovisual media is ``inherently complex, combining multiple information modalities over time'' in ways that introduce ``additional cognitive and language barriers that cannot be addressed through translating information from one mode to another.'' Their work demonstrates that personalization can be taken ``to its logical next step---creating bespoke interventions for a single individual that address their specific access needs'' \cite{nevskyetal25}. This approach aligns with Harper's \cite{harper2007there} argument for ``design-for-one'' that accommodates ``every person's uniqueness through providing an opportunity to change the viewing paradigm through highly personalised content adaptation'' \cite{nevskyetal25}. The concept of ``content domestication''---adapting media to meet the viewer's idiosyncrasies from the outset \cite{nevskyetal25}---resonates directly with our participants' desires for platforms that adapt to their sensory needs rather than requiring constant manual intervention.

\subsection{Participatory Design with Disabled and Neurodivergent Communities}

Critical disability studies scholarship emphasizes "nothing about us without us"â€”the imperative to include disabled people in designing for accessibility \cite{spiel2020nothing}. Bennett and Rosner \cite{bennett2019promise} critique empathy-driven design approaches, arguing that empathy cannot substitute for lived expertise and participatory engagement. Frauenberger \cite{frauenberger2015disability} advocates for disability-led design that recognizes disabled people as experts in their own experiences. This perspective aligns with the neurodiversity movement within disability studies, which advocates for acknowledging neurological differences as natural human variations rather than deficits to be corrected \cite{sabinson2025blowfish}. Sabinson \cite{sabinson2024pictorial} elaborates this stance through examination of behavior intervention designs, noting that such designs ``often mirror techniques utilized in behavioral therapies aimed at preserving neurotypical social expectations at the expense of individual autonomy.'' Drawing on the Behavior Design Lab's ethical framework, which recommends that designers ``focus their research and efforts on positive change and helping people succeed and feel successful at doing what they already want to do,'' Sabinson argues for ``more objects, environments, and systems that support innate and self-determined behaviors, and more caution for products and technologies that aim to change behaviors. In other words, fewer fixes'' \cite{sabinson2024pictorial}.

In practice, participatory design with disabled communities faces challenges including power imbalances \cite{spiel_et_al_22,sum_et_al_22}, tokenistic participation, and inaccessible design processes themselves \cite{stephens2023accessibility}. Spiel et al. \cite{spiel2019agency} document how autistic children's agency is often constrained in technology research. Ringland and Wolf \cite{ringland_et_al_21} reflect on creating accessible social spaces, emphasizing ongoing negotiation rather than fixed solutions. To ensure user research is inclusive, it is essential to accommodate neurodivergent participants who may experience anxiety or sensory sensitivities. Environmental factors such as intense lighting, excessive noise, or crowded spaces can cause significant distress, potentially leading participants to withdraw from the study entirely \cite{lukava2022two}. Li et al. \cite{li2024codesigning} demonstrate how co-design with people who stutter can address gaps in inclusive video conferencing, revealing that ``individuals with invisible disabilities, such as those who stutter, encounter these obstacles with extra hurdles as current VC technologies are often designed with assumptions and heuristics about the user's speech and speech behaviors that are incompatible with stuttering.'' Their work finds that ``the result is functional inaccessibility and emotional harm to users who stutter'' \cite{li2024codesigning}, underscoring the need to center lived expertise in design processes.

Autobiographical design offers another promising approach. Sabinson \cite{sabinson2025blowfish} explores how designing for one's own sensory needs can reveal new possibilities for ethical behavior change in HCI, demonstrating that ``sensory needs are often suppressed in order to conform to social norms'' and that centering neurodivergent autonomy---rather than imposing change on others---opens new design pathways.

Emerging work demonstrates how bespoke co-design can address the limitations of population-level accessibility interventions. Nevsky et al. \cite{nevskyetal25} argue that while co-designing for a population ``we miss the core goal of personalisation---the ability to support every individual, no matter how complex their needs or the lack of scalability.'' Their work with people with aphasia required adapting standard participatory design methods, including ``minimising language-based tasks in favour of using alternative communication,'' having ``short and direct tasks introduced verbally by the researchers,'' and ``incorporating tangible props to represent interactions or concepts'' \cite{nevskyetal25}. These methodological adaptations resonate with our approach to engaging sensory-sensitive participants who may experience distress when discussing triggering technologies.

Our methodological contribution builds on this foundation by documenting how we engaged sensory-sensitive participants in design exploration. We reflect on challenges including the contested nature of misophonia as a diagnosis, power dynamics between researchers and participants seeking validation, and the difficulty of participatory design when participants experience distress from the very technologies we're discussing.

% Combined Critical Perspectives & Technology Non-Use Section
% All original citations preserved

\subsection{Platform Affordances and Sensory Exclusion}

Platforms shape social behavior through affordances---the actions they enable, constrain, or discourage \cite{Fuchsberger_et_al_13_materiality}. Recent platform studies work examines how design choices embed values and distribute power \cite{light2018walkthrough,eglash2006technology}, while Becker et al. \cite{becker2019audiovisual} argue for examining the ``intersection between Human-Computer Interaction Studies and Media Studies,'' noting that integrated audiovisual content has fundamentally altered how ``audiovisual products act on the perception of audiences.'' Im et al. \cite{Im_um_2021} propose affirmative consent as a design principle, arguing that users should actively opt-in rather than passively accept defaults. We extend this critical lens to audiovisual affordances, asking: whose sensory experiences are centered as ``default,'' and whose are marked as exceptions requiring accommodation? When platforms like TikTok auto-play audio or Zoom encourages cameras-on participation, they enact assumptions about sensory tolerance that exclude sensory-sensitive users.

\subsubsection{Technology Non-Use as Exclusion by Design}

HCI scholarship increasingly recognizes that understanding refusal, limiting, or abandonment is as important as studying adoption. Baumer et al. introduced categories like ``lagging resistance'' and ``symbolic non-use'' to explain deliberate disengagement \cite{baumer_et_al_13}, framing non-use as meaningful behavior through terms like ``resisters'' and ``liminal users'' \cite{baumer_et_al_14}. Densmore examined when technologies should fail \cite{densmore2012}, while Barocas et al. frame refusal as ethical responsibility \cite{barocas2020}. Building on DiSalvo's conception of design as space for contestation \cite{disalvo2015adversarial}, Zong and Matias articulate ``data refusal from below'' centering resistance to algorithmic systems \cite{Zong_Matias_24}, and Garcia et al. position refusal as feminist practice \cite{garciaetal2020}.

However, for many of our participants, ``choice'' overstates their agency---they are excluded by design that does not account for their needs. Technology non-use research \cite{Fuchsberger_et_al_14} has examined why people choose not to engage with technologies, yet this framing can obscure structural exclusion. We reframe ``non-use'' as exclusion by design, shifting responsibility from users who ``can't handle'' platforms to designers who have not built for sensory diversity.

\subsubsection{Relational and Contextual Disengagement}

Disengagement from digital systems is fundamentally a product of specific social contexts and relationships. Light and Akama argue that for designers working with communities facing structural marginalization, "emotional attunement" is vital, as digital habits are deeply embedded in collective histories and emotional bonds \cite{light_akama_2012}.

Recent empirical studies illustrate how this refusal manifests across different environments:

\begin{itemize}
    \item \textbf{Negotiated Representation:} Queer individuals often navigate the complexities of identity by resisting the representation dynamics found in targeted advertising \cite{sampsondenae23}.
    \item \textbf{Systemic Rejection:} Research has characterized the total abandonment of systems \cite{johnson1989mental} and the specific refusal to adopt AI within professional workplaces \cite{chatetal25}.
    \item \textbf{Participatory Refusal:} Within design research, Robinson et al. suggest that refusing to engage can itself be a form of participation \cite{robinsonetal2020}, while Dourish et al. point to the heavy emotional toll placed on communities that are frequently the subject of academic study \cite{dourishetal2020}.
\end{itemize}
 
The methods people use to avoid or limit technology vary significantly based on their unique circumstances and demographics.

Adaptive Tactics
For many, non-use is a proactive choice to maintain well-being or control. Brewer et al. found that older adults frequently opt for phone calls over online forums to avoid social friction \cite{brewer_et_al_21}. Similarly, some users practice "functional minimalism," a strategy of strictly selecting only the digital tools they find manageable \cite{sheehan_23}. For older migrants, maintaining agency often involves a reliance on analog methods or family support systems rather than digital-first solutions \cite{zhao_et_al_23}.

Often, disengagement is forced by exclusionary design rather than personal choice:

\begin{itemize}
    \item \textbf{Language \& Translation:} Linguistic obstacles frequently require older users to depend on family members for digital translation \cite{wong-villacres_et_al_19, yuan_et_al_24}.
    \item \textbf{Inaccessible Services}: Healthcare portals that prioritize English force patients to rely heavily on caregivers for basic access \cite{le_et_al_24}.
    \item \textbf{Algorithmic Bias:} Voice-activated AI that fails to accurately process accented speech often leads users to abandon the technology entirely \cite{mittal_et_al_25}.
\end{itemize}

Ultimately, these behaviors demonstrate that "non-use" is rarely a reflection of a user's lack of skill; rather, it is a response to systemic design exclusions that fail to account for diverse human experiences.

\subsubsection{Neurodivergent Users and Platform Friction}

Research on neurodivergent people's social media use reveals both affordances for connection and barriers to participation. Jiang et al.'s \cite{jiang2025shifting} systematic literature review finds that research on neurodivergent people and social computing is ``largely medicalized, adheres to historical stereotypes, and is insensitive to the wide spectrum of neurodivergent identities.'' They note that neurodivergent people ``often rely on social technologies for a wider variety of tasks compared to neurotypical users, including education, information, and emotional support,'' yet face ``values for social experience encoded by social technologies'' that may conflict with their needs \cite{jiang2025shifting}. Participants with ADHD were wary of platforms that ``enraptured people through their addicting algorithms and dark patterns,'' describing TikTok as synonymous with addiction---``a dopamine slot machine''---with some explicitly avoiding the application \cite{jiang2025shifting}.

Yet platform design creates friction for neurodivergent users in multiple ways. Algorithmic content curation surfaces unpredictable sensory experiences. Auto-play features remove user control. Social norms around ``authentic'' content---like mukbang videos or ASMR---become unavoidable triggers. Our work contributes by centering design solutions rather than documentation of challenges, proposing specific interventions that would allow platforms to support diverse sensory needs without sacrificing social connection.

Our work contributes by centering design solutions rather than documentation of challenges, proposing specific interventions that would allow platforms to support diverse sensory needs without sacrificing social connection.

\section{Methods}

Due to the complicated nature of discourse around a contested condition, we chose a
qualitative research design. Qualitative interviews are widely used in social science to explore
individuals’ experiences and perceptions of a phenomenon. Semi-structured interviews offer a
balance between the flexibility of an open-ended conversation and the structure needed to ensure
that key topics are covered. It is imperative in providing “some degree of standardization to ensure
key issues are covered, while retaining the flexibility to allow the interview to naturally flow and
for participants to introduce their own perspectives” \cite{clarke2013successful}. In our study, we employed semi-
structured qualitative interviews to explore the experiences of individuals with misophonia. Our
interview guide was designed to elicit participants’ personal experiences of living with
misophonia, their strategies for coping with it, and their offline and online interactions discussing
it, as well as their discussions with healthcare professionals. Using open-ended questions and
probes, we aimed to gather rich and detailed data on the subjective experience of misophonia from
the participants’ own perspectives.

\subsection{Data collection and screening questionnaire}
To find out how individuals with misophonia find support and engage in spaces online, we
interviewed 16 users. Prior to commencing interviews, we received approval from our Institutional
Review Board. We recruited participants by advertising in online spaces on
three social media platforms: Reddit subreddits (n = 4), Instagram (n = 4), and Facebook (n = 10).
We asked moderators and page owners for permission to post the recruiting message, which
included a link to a Qualtrics screening survey and consent form. A poster advertising the study
was posted in the Facebook group. Each participant was informed in writing of their recruitment
to ensure they were aware of their rights and what information would be collected from them.

During the interview, each participant was reminded again of their rights and the voluntary nature
of their participation. All collected data, including recordings or transcripts of the interview, was
de-identified and destroyed after data analysis. No participants under the age of consent were
recruited to this study.

Participants for this study were recruited based on multiple factors. The first criterion was
having the disorder misophonia or having family members who have the disorder. The second
criterion was being part of one or more online spaces dedicated to misophonia in some way. The
final criterion was that participants must reside in the United States and be at least 18 years of age
or older. Once ensured that the participant met all criteria, they were contacted to schedule and
email over Zoom or telephone. We received a total of 76 questionnaire responses and filtered out individuals who did not
meet the study criteria. An invitation to join the study was sent to n = 27 candidates in total, with
a final number of participants being n = 16. The average demographics and details of the
participants are summarized in Table \ref{tab:participants}. Of our participants, 81.3\% identified as female, with 18.8\% identifying
as male. The respondent race demographics are: 6.3\% Asian and 93.8\% White. Participants represented 12 states across the United States.  

\begin{table}[!ht]
    \centering
    \caption{\label{tab:participants} This table lists our participants along with their self-reported gender and misophonia triggers.}
    \begin{tabular}{|l|p{9cm}|l|}
    \hline
        Pseudonym  & Triggers  & Gender   \\ \hline
        Evelyn  & Chewing, slurping, tongue clicking, sniffling, tapping, joint cracking, water dripping, pen clicking, shoes scuffing, snoring, glasses/silverware scraping  & F   \\ \hline
        Nadia  & Chewing, slurping, tongue clicking, sniffling, tapping, joint cracking, water dripping, shoes scuffing, snoring, glasses/silverware scraping, clocks  & F   \\ \hline
        Kathleen  & Chewing, slurping, tongue clicking, sniffling, tapping, snoring, glasses/silverware scraping, hands/feet rubbing together  & F   \\ \hline
        Kendra  & Chewing, slurping, tongue clicking, nail clipping, glasses/silverware scraping, keyboard typing  & F   \\ \hline
        Nathan  & Chewing, slurping, tongue clicking, nail clipping, pen clicking  & M   \\ \hline
        Kristie  & Chewing, slurping, tongue clicking, sniffling, tapping, joint cracking, pen clicking, snoring, glassware/silverware scraping  & F   \\ \hline
        Vlad  & Chewing, slurping, tongue clicking, sniffling, snoring  & M   \\ \hline
        Rhys  & Chewing, slurping, tongue clicking  & F   \\ \hline
        Amelia  & Chewing, slurping, tongue clicking, tapping, water dripping, pen clicking, snoring, glasses/silverware scraping, rubbing hands together  & F   \\ \hline
        Lydia  & Chewing, slurping, tongue clicking, sniffling, dry mouth noises  & F   \\ \hline
        Chuck  & Chewing, slurping, tongue clicking, sniffling, tapping, joint cracking, nail clipping, water dripping, pen clicking, shoes scuffing, snoring, glasses/silverware scraping  & M   \\ \hline
        Erica  & Chewing, slurping, tongue clicking, sniffling, tapping, water dripping, pen clicking, snoring, glasses/silverware scraping  & F   \\ \hline
        Lola  & Chewing, slurping, tongue clicking, sniffling, joint cracking, nail clipping, pen clicking, shoes scuffing, glasses/silverware scraping  & F   \\ \hline
        Ashley  & Chewing, slurping, tongue clicking, sniffling, tapping, nail clipping, shoes scuffing, glasses/silverware scraping, throat clearing  & F   \\ \hline
        Cindy  & Chewing, slurping, tongue clicking, sniffling, tapping, pen clicking, shoes scuffing, snoring, glasses/silverware scraping, pet noises, swallowing, scratching, throat clearing  & F   \\ \hline
        Justine  & Chewing, slurping, tongue clicking, sniffling, tapping, pen clicking  & F  \\ \hline
    \end{tabular}
    
\end{table}

\subsection{Interview Design}
We developed the interview protocol  to
allow participants freedom in their answers while ensuring consistent responses for analysis. We
built a set of interview questions with follow-up probing questions. Interviews were conducted
until data saturation was reached. 

Our semi-structured protocol covered:
\begin{itemize}
\item Lived experiences of misophonia (triggers, contexts, impacts)
\item Current coping strategies (technological and social)
\item Technology use and avoidance patterns across platforms
\item Experiences with seeking recognition/accommodation
\item Positive experiences and effective supports
\item Design ideas and imagined solutions
\end{itemize}

\subsection{Analytical approach}
All interviews were recorded and transcribed using Otter AI. During transcription, all identifying details were replaced with pseudonyms to ensure confidentiality. The resulting transcripts were imported into NVivo for qualitative analysis, where data were systematically coded to identify and examine emerging themes.

The analysis followed a combination of open and axial coding. In the open coding phase, interview segments were broken down into smaller units to surface key concepts and relationships. Axial coding then built on these insights by linking categories and refining connections between them, a process that breaks raw data into units, uncovers new concepts and relationships, and systematically develops categories that are then recombined to build theory \cite{wall1999sentence}. This approach enabled deeper exploration of relationships across participants’ experiences and was therefore selected as the primary coding strategy.

The thematic analysis was carried out independently using the established codebook to ensure consistency and rigor throughout the process.

\section{Findings}
We present our findings as design insights that reveal gaps in current platform affordances and point toward design solutions. We organize findings around three key contexts where sensory accessibility failures emerged most strongly.

\subsection{Seeking Recognition: The Design of Medical and Social Legitimacy Channels}
Misophonia is not well understood or recognized in the medical field, let alone by friends
and family of those who suffer from it. The lack of recognition can be discouraging for those
experiencing the disorder, as they may face dismissal or disbelief from professionals. In this excerpt, Evelyn describes how the obscurity of
the disorder makes her feel unimportant and despondent.
\begin{quote}
``Like you've heard of other
things like anxiety, bipolar, schizophrenia, but like, this is not out in the open. This is not something that people ever talk about. So it's like, ‘Are you lying?’ [...] It
kind of makes you feel unimportant.''
\end{quote}

Kathleen expressed a similar exhaustion in attempting to find mental health professionals who
understand and can attempt to treat misophonia. She says that “There are no mental health
professionals that I can find that have any sort of understanding of like, what it is or how to treat
it. And it is exhausting.”

Rhys highlighted the difficulty in diagnosing misophonia, as it is not defined by a clear set
of criteria or is included in the current Diagnostic Statistical Manual of Mental Disorders (DSM).
Despite the lack of ability to be diagnosed by a mental health professional, the disorder can
significantly impact an individual's daily life. This lack of recognition can also lead to self-doubt
and feelings of isolation and inadequacy, as those with misophonia may feel as though they are
the only ones experiencing their symptoms. This was a common theme among participants and
characterized their drive to find communities online with others suffering from the disorder with
whom they can relate socially.

Kendra found validation in
hearing from a professional that her symptoms were related to misophonia. She also expressed
interest in obtaining an official diagnosis in order to receive accommodations at work through the
Americans with Disabilities Act (ADA). Kristie similarly noted the challenge of finding
accommodations for misophonia, as it is not currently classified in the DSM, suggesting the
importance of advocating for its inclusion in the future so as to better support individuals with
misophonia in schools and the workplace. 

\fbox{%
  \parbox{\linewidth}{%
\textbf{Design Implication:} Platforms become critical sites for validation and information-sharing when medical systems fail. Yet if platforms simultaneously exclude users through hostile sensory design, they create a cruel double bind: seek community online while being sensorially assaulted by the platforms that host it.
  }%
}

However, some misophones described how they had supportive experiences with their partners. 
For example, Amelia's friend and former roommate in college understood what she was going through. This built-in support system has been a source of comfort for Amelia, as she has always been surrounded by misophonic triggers. Echoing Amerlia, many participants, such as Evelyn, Justine,
Nathan, and Kathleen have acknowledged that their partners have been supportive and
understanding, often putting in the effort to research and help them cope with their triggers. Positive experiences have not just been a part of home-life; Chuck had a positive experience at his prior job, where he discussed how his supervisor had previous knowledge about
misophonia and made accommodations in the workplace to minimize trigger sounds. He explains
the importance of his boss’ understanding and attempt to make accommodations in the following
excerpt.
\begin{quote}
    “They weren't perfect. But there was a no food and drink or no food and gum policy
implemented in the office. So people would have to step out in the hall or, or
we had a dining area on our floor. And they actually bought me some noise canceling headphones that I used. So I thought that was
pretty aboveboard, I think, from talking to other people, that was actually an
unusual experience. A lot of a lot of people don't get those, even though they're
federally required to have reasonable accommodations. A lot of people just get
denial and pushback and, and minimization and stuff and don't actually end up
getting those.”
\end{quote}

\subsubsection{Disclosure Risks: When Advocacy Backfires}

While some participants found supportive responses when disclosing their misophonia, others experienced deliberate antagonism. Lydia described a workplace situation where disclosure led to intentional triggering:

\begin{quote}
``I ended up telling her I was like, listen, like, here's what I had this thing and can you just like, not eat at the desk, you know, please. And then she pretty much just, like, amped it up after that and would like look at me and like, pop these like tomato cherries in her mouth while she was looking at me...it's like, you know, like, so, not to say that, that kind of like, you know, traumatize me a little bit? Honestly. It's like, I don't want to tell the wrong person.''
\end{quote}

This experience created lasting hesitation about future disclosure. The risk of disclosure backfiring---having one's vulnerability weaponized---represents a significant barrier to seeking accommodation. Family members also sometimes responded with mockery rather than support. Multiple participants described relatives who would ``poke fun'' at their condition, deliberately making triggering sounds after learning about the condition.

\subsection{The Closest Relationships as Greatest Triggers}

\subsubsection{Moderating closed misophonia communities online}
Chuck notes that one solution to decrease such occurrences which he also faced can be the creation of private subreddits in which any users have to explicitly gain access by satisfying questions addressed to them by moderators who act as gatekeepers. However, he concludes that ``there's probably some people who would like to be in those groups...[but] don't really want to apply to join the group because they do not know what's in it.'' In other words, while this setup might reduce unwanted negative behavior, it will also limit the number of people who can get support in the community. Case in point, Lola says that ``a private Facebook group or is that the one where like you apply and get accepted [by moderators...] Yeah, it's definitely a barrier, would you be more open to posting if it was not private? Like if it was easier to join this communities? Maybe? […] If it was on like, Instagram, I feel like it'd be more accessible but Facebook, it just not as accessible for me personally.'' 

Accessibility is not the only danger of relying on a select few gatekeepers. Ashley describes how, after her bad experiences on Reddit, she decided to join a closed Facebook support group which was moderated by a physician focusing on misophonia. While she initially thought the community was a great source of information which is verified by a medical professional, and where there can be open discussions between misophones and medical professionals in the field. Ashley said that she hadn't ``really participated much in [the Facebook group] lately...because there's no longer a moderator.'' Having a small number of moderators (in this case, a single moderator) made this community less versatile over time. Ashley had another bad experience where the moderator of an online group presented what she saw as misinformation. Specifically, the group moderator, a medical professional, identified misophonia as a audiological disorder which does not have any psychological or neurological dimentions. This in fact, contradicts most scholarship on misophonia - introduced at the top of this paper (c.f.\cite{swedo2022consensus,cavanna2015misophonia,potgieter2019misophonia,jaswal2021misokinesia,schroder2013misophonia,taylor2017misophonia}). When Ashely introduced studies similar to these presented here in this closed online community she was
\begin{quote}
     ``kicked out of that group, because I kept saying, I kept insisting that this is not a hearing problem...No, this is not a hearing problem. Because I had already recognized that I had a visual trigger. It was very clear to me that there was a psychological component. And she outright rejected the idea that it could have anything to do with psychology.''
\end{quote}
At the time Ashley was interviewed, she was removed from one Facebook group, and ceased posting to Reddit and another Facebook group. This is one example of technology non-use. Below, we explore other examples of technology non-use presented in our findings. 

The implications for platform design are significant: moderation systems that concentrate definitional authority in single gatekeepers risk replicating the dismissal and delegitimization that participants experienced in clinical settings. A more \textit{communitarian} approach to moderation---one that distributes epistemic authority across community members and privileges lived experience alongside professional expertise---could reduce epistemic trauma while maintaining community quality. This might include rotating moderation responsibilities, requiring multiple moderator consensus for membership decisions, or creating explicit pathways for members to contest moderation decisions that dismiss their experiential knowledge. 

\fbox{%
  \parbox{\linewidth}{%
    \textbf{Design implication:} Just as supportive individuals make accommodations, platforms could build accommodation into affordances rather than relying on disclosure and negotiation. Design can embody the understanding that some users need partners and supervisors to demonstrate individually. \textit{$\rightarrow$ See \S\ref{sec:design-collaborative}: Support Collaborative Boundary Negotiation.}
  }%
}

\subsection{Algorithmic Assault: Sensory Triggers in Social Media Feeds}
Participants identified social media platform -- particularly TikTok and Instagram -- as particularly problematic due to audiovisual affordances that prioritize engagement over user control.

\subsubsection{ASMR and the Politics of "Relaxing" Content}

TikTok's algorithm heavily promoted ASMR (Autonomous Sensory Meridian Response) content during our study period (2022-2023). ASMR videos feature deliberate triggering sounds -- tapping, scratching, whispering, mouth sounds -- intended to create pleasurable tingling sensations for some viewers \cite{poerio2016could}. For misophonic users, these same sounds trigger distress.

Kathleen described a "love-hate relationship" with TikTok: \textit{"The some of the sounds on there are just so bad and like, ASMR No, thank you. And ASMR blew up and everybody was doing it, and I just like, I cannot stand it. And so there were certain apps like Tiktok or even like Instagram...I found myself avoiding those because all of the content was just like unbearable to listen to, or if I was on them, I was just always needed like could not have the sound up."}

Using TikTok with sound muted defeats the platform's core value proposition -- its audio-synchronized content. Cindy stopped using TikTok entirely because she "could not watch the videos without sound."

\fbox{%
  \parbox{\linewidth}{%
    \textbf{Design Implication:} TikTok's design assumes audio is universally desirable, with no option to filter content by sensory properties or set audio preferences. The "For You" algorithm optimizes for engagement without considering that engagement metrics may mask exclusion -- users might scroll quickly past triggering content, appearing "engaged" while experiencing distress.

    Platforms need content-level audio controls that let users filter by audio characteristics or genres (e.g., "reduce ASMR content," "filter eating sounds"). This requires treating sensory accessibility as a content moderation and curation issue, not just an interface design issue. \textit{$\rightarrow$ See \S\ref{sec:design-filtering}: Disaggregate Audiovisual Control and \S\ref{sec:design-predictability}: Design for Sensory Predictability.}
  }%
}

\subsubsection{Platform Audio Features: Inadequate Controls for Sensory Management}

Participants identified specific design limitations in how platforms handle audio content. While TikTok technically allows users to block specific audio clips, participants noted this feature was woefully inadequate for managing triggers. Popular sounds are frequently re-uploaded by different users, meaning blocking one instance of a triggering audio does not prevent exposure to duplicates. A user must block \textit{all} instances of a particular sound individually---an impossible task given the volume of content.

Cindy articulated the fundamental incompatibility between TikTok's audio-centric design and her sensory needs: she ``could not watch the videos without sound,'' yet watching with sound meant constant exposure to potential triggers. This created a binary choice between meaningful platform engagement and sensory safety---a choice that ultimately led her to abandon the platform entirely.

Participants suggested that platforms could implement trigger warnings similar to those used for violent or sensitive content, but specifically for auditory properties. The absence of such warnings meant participants had no way to anticipate triggering content before exposure. Unlike visual content warnings that can be read before viewing, audio triggers often occur immediately upon content loading, providing no buffer for preparation or avoidance.

\fbox{%
  \parbox{\linewidth}{%
    \textbf{Design Implication:} Current audio blocking mechanisms are reactive rather than proactive, requiring users to encounter triggers before blocking them. Platforms should implement: (1) category-level audio blocking (e.g., ``block all ASMR-tagged content'') rather than instance-by-instance blocking; (2) audio preview features that allow users to assess content before full exposure; and (3) sensory content warnings that appear before audio plays, not after. \textit{$\rightarrow$ See \S\ref{sec:design-predictability}: Design for Sensory Predictability.}
  }%
}

\subsubsection{Secondhand Scrolling: Ambient Technology Use as Trigger}

Nadia described being triggered not by her own social media use but by overhearing others' use: \textit{"I'll be Like in a room with a friend and they be like scrolling through TikTok and I like wouldn't be able to fully hear it, but I could still hear it. And it just be like the switching of sound so quickly...That became my trigger. And I'd be like, every one of my friends knows that they're going to like watch TikTok around me they have to put in headphones."}

A striking pattern emerged across interviews: participants consistently reported that triggers were most intense with family members and loved ones, particularly parental figures. Evelyn described the paradox: \textit{``my husband's breathing will bother me. So it's like, I cannot tell him to stop breathing.''} Nathan similarly noted that \textit{``the first person that I remember bothering me about it, you know, that it bothered me was my mother...it seems like it's the people that are closest.''}

Lola's therapist provided one framework for understanding this pattern: \textit{``the closer you get with someone, and the more time you spend with someone is like, really, it there's an emotional aspect with misophonia where, like, if you're angry at them, or you have some type of emotion towards them, you're more likely to get triggered at them.''} This emotional component intensified the guilt participants felt, creating what several described as a ``double burden''---experiencing physiological distress while simultaneously feeling guilty for that response toward people they love.

The domestic context proved particularly challenging because triggers occurred in spaces meant to feel safe. Amelia noted: \textit{``it was at home with my husband where I wanted, you know where you should be the happiest and safest. And I think what, like, I got to be quiet so he doesn't hear me say this, but...''} The impossibility of escaping triggers in one's own home led some participants to drastic measures. Lola ultimately \textit{``moved into this apartment literally, like last week''} specifically to escape roommate triggers, while Amelia's family eventually asked her father-in-law to move out, partly due to the sustained sensory distress his presence caused.

\fbox{%
  \parbox{\linewidth}{%
    \textbf{Design Implication:} Sensory accessibility extends beyond individual user-platform interactions to encompass shared physical and social contexts. Platform use rarely occurs in isolation---it happens in living rooms, offices, and public spaces where sensory boundaries blur and others' technology use can introduce triggers into previously safe environments. Design solutions must therefore consider broader sensory ecologies rather than user-platform dyads alone. Smart home integration for household-wide audio preferences, notification systems for sensory-sensitive periods, and platform-level defaults that reduce the need for vulnerable individual disclosure can achieve accommodation without exposing users to antagonistic responses or retaliation. \textit{$\rightarrow$ See \S\ref{sec:design-collaborative}: Support Collaborative Boundary Negotiation.}
  }%
}

This points toward features like "public space mode" with reduced audio by default, or social features that let users share sensory preferences (e.g., "I'm sensitive to audio bleed -- please use headphones"). It also raises questions about platform responsibility for shaping social norms around considerate use.

\subsection{Remote Work, Remote Triggers: Video Conferencing Platforms}

The shift to remote work and online socialization during COVID-19 made video conferencing unavoidable. Yet Zoom, Discord, and similar platforms revealed critical gaps in audiovisual control.

\subsubsection{Eating on Camera: Norms, Necessity, and the Missing Middle}

Multiple participants described distress from colleagues eating during video calls. Nathan specifically mentioned eating as a trigger during Zoom meetings. Critically, participants emphasized that triggers were not purely auditory---even when audio was muted, the \textit{visual} of someone eating could trigger misophonic responses. Amelia described confronting a music teacher who ate during lessons: \textit{"I felt weird telling her this because that's she's,she's making she thinks it's funny...I'm like, no, no, this is probably like an actual thing...some of [your students] might be able to hear you, or even just the the visual of it. Like it's actually causing them pain on the inside, like mental pain and anguish."}

Rhys characterized Zoom as ``so auditory'' in nature, highlighting how the platform's design centered audio in ways that amplified trigger exposure. She noted that video calls created an environment where triggers were inescapable---unlike in-person settings where one might physically distance from a trigger source, video conferencing compressed all participants into the same auditory space.

Participants noted that eating on camera became normalized on Zoom in ways it wouldn't be in-person. Long meetings spanning meal times made eating seem necessary. Yet for misophonic users, this created an impossible choice: advocate for limiting others' behavior (feeling like "Debbie Downer" as Amelia put it), mute everyone and miss the meeting, or endure physiological distress.

The burden of disclosure weighed heavily on participants. Ashley described the emotional labor of repeatedly explaining her needs:
\begin{quote}
``I feel like that's just an extra burden on them. And I always get, like, a sigh when I say what you're doing right now really bothers me. I'll get an exasperated response, and that really hurts me. So as much as they try to accept and accommodate. Maybe it's asking too much, maybe it's impossible.''
\end{quote}

This sense of being burdensome led many participants to suffer in silence rather than request accommodations, particularly in professional contexts where they feared being perceived as difficult or unreasonable.

\fbox{%
  \parbox{\linewidth}{%
    \textbf{Design Implication:}  Zoom's binary audio options -- mute everyone or hear everyone -- fail to account for selective sensory management needs. The platform assumes either full participation or full withdrawal, with no middle ground.

    Users need individual-level audio control: ability to lower (not mute) specific participants, apply audio filtering to individual streams, or temporarily mute someone without losing visual presence indicators. This preserves participation while reducing sensory distress. \textit{$\rightarrow$ See \S\ref{sec:design-filtering}: Disaggregate Audiovisual Control.}
  }%
}

\subsubsection{Gaming and Gathering: Long Sessions and Accumulated Distress}

The shift to remote socialization during the pandemic meant that leisure activities---not just work---moved online, often for extended periods. Rhys described how remote social activities amplified challenges: \textit{"I play Dungeons and Dragons with my friends online a lot [using Discord]. And it lasts like, three hours long, right? And you cannot expect it cannot be like, and I shall dictate that none of you will eat dinner."}

The duration of these sessions created cumulative exposure to triggers. Unlike a brief meeting where one might endure temporary discomfort, multi-hour gaming sessions meant sustained stress. Participants could not simply ``power through'' a three-hour session---the physiological toll was too great.

Discord offered more granular controls than Zoom---Rhys ``discovered the very important function of just sort of being able to like, like, lower certain people's volumes.'' However, even these controls had limitations: reducing someone's volume made them harder to hear when they were speaking meaningfully, not just making trigger sounds. This created a cruel trade-off: reduce trigger exposure but miss important contributions to the game, or maintain full audio and suffer.

Participants also noted the social complexity of managing triggers in leisure contexts. In work settings, one might invoke professional norms or ADA accommodations. Among friends, the dynamic was different---requests to modify behavior felt like impositions on intimacy. Evelyn captured this tension: ``my husband's breathing will bother me. So it's like, I cannot tell him to stop breathing.'' The impossibility of asking loved ones to cease fundamental human activities created profound guilt.

\fbox{%
  \parbox{\linewidth}{%
    \textbf{Design Implication:} Current volume controls are too coarse-grained. Users need the ability to reduce ambient or incidental sounds from a participant (eating, tapping) without reducing their speech volume equally.

    This points toward audio filtering at the source -- detecting and reducing specific sound types (eating, tapping, keyboard) while preserving speech. Technical approaches might include trained audio classifiers that let users say "reduce eating sounds from this person by 80\% but keep their voice at 100\%." \textit{$\rightarrow$ See \S\ref{sec:design-realtime}: Enable Real-Time Sensory Filtering.}
  }%
}

\subsubsection{Visual Triggers and Misokinesia: The Overlooked Dimension}

While auditory triggers dominated discussions, several participants described visual triggers that platforms similarly failed to accommodate. Amelia noted: \textit{``my husband bounces his knee. He's one of those people who can't. And I have to I have to just put my hand on his knee. And sometimes it's not even I see it, but like I can feel the table shaking.''} This illustrates how misokinesia (visual trigger sensitivity) often co-occurs with misophonia.

Nathan explicitly described the visual component of eating triggers on Zoom: \textit{``I'm not fond of people eating on Zoom...I don't want to hear it. And I don't want to watch it...It's not as bad. I mean, you know, if hearing them is maybe a nine, you know, seeing it as, like a two or three.''} While less intense than auditory triggers, visual triggers still caused distress and were equally uncontrollable on current platforms.

Kristie described visual triggers extending to seemingly minor movements: \textit{``my husband...he'll scratch his balls. And I'm like, Are you like a monkey?...it triggers me. That's one of the biggest triggers.''} She also noted the impossibility of fully escaping visual triggers: \textit{``Maybe I need one of those things like, you know, you see horses have those things where it's like blinders like that so that I don't see what people do.''}

\fbox{%
  \parbox{\linewidth}{%
    \textbf{Design Implication:} Platform accessibility features should address visual as well as auditory triggers. Video conferencing platforms could offer per-participant video blur options, the ability to reduce or hide specific video feeds while maintaining audio, or automatic detection and blurring of common visual triggers like eating or repetitive movements. Social media platforms could extend content warnings to include visual trigger categories. \textit{$\rightarrow$ See \S\ref{sec:design-visual}: Address Visual Triggers Through Selective Video Management.}
  }%
}

\section{Discussion}
In this study, we investigated how misophonia shapes technology use and non-use, the coping strategies individuals employ to manage technology-mediated trigger exposure, and the design principles that could guide the development of sensory-accessible platforms. Our findings reveal patterns of technology-mediated exclusion that echo and extend challenges documented across the neurodivergent community.

\subsection{Patterns of Technology Use and Non-Use (RQ1)}

Our first research question asked how misophonia shapes patterns of technology use and non-use across social media, video conferencing, and leisure platforms. We found that audiovisual features of social media sites like Instagram and TikTok were prime triggers for misophones because they could not control triggers by muting videos. Additionally, we found that the use of these technologies by friends and family might itself be triggering if they were within earshot, especially when they are listening to triggering audio (e.g., ASMR). This ``secondhand'' triggering through others' technology use highlights how sensory accessibility extends beyond individual user-platform interactions to encompass broader sensory ecologies of shared physical and digital spaces.

The patterns of technology use and non-use we observed align with theoretical frameworks of selective engagement and partial domestication \cite{salovaara2011everyday,sorensen2006domestication}. Participants did not simply adopt or reject technologies wholesale; rather, they engaged in nuanced negotiations with platform affordances. Some participants maintained presence on platforms like TikTok but with carefully curated feeds and strict muting practices. Others abandoned certain platforms entirely after repeated trigger exposure made continued use untenable. This spectrum of engagement---from selective use to episodic engagement to complete abandonment---reflects what Gorm and Shklovski \cite{gorm2019episodic} describe as ``self-care'' through technology disengagement, where users ``pause'' domestication when constant monitoring produces anxiety.

Everyday activities like eating while using services like Discord and Zoom triggered misophones, which made any communal experiences (whether for work or leisure) mediated by these technologies difficult to cope with. Li et al. \cite{li2024codesigning} found similar challenges for people who stutter, noting that ``VC technologies are often designed with assumptions and heuristics about the user's speech and speech behaviors that are incompatible'' with diverse needs. The parallel is instructive: just as voice interfaces assume fluent speech, video platforms assume uniform sensory tolerance. Both represent what Li et al. call the ``functional inaccessibility and emotional harm'' that results when technologies embed normative assumptions about user capabilities.

The broader implications of our findings resonate with earlier observations that neurodivergent people are often positioned as consumers rather than active shapers of digital environments, with ``simplistic and restrictive systems that prevent users from `being social' in a way that feels natural and enjoyable'' \cite{baillargeon2025social}. By reframing technology non-use as ``exclusion by design'' rather than user deficiency, we shift analytical and practical responsibility from individuals who ``can't handle'' platforms to designers who haven't built for sensory diversity. This perspective aligns with Sabinson's \cite{sabinson2024pictorial} argument for ``supporting self-determined behaviors instead of aiming to `fix' behaviors deemed abnormal or problematic by society''---designing for what users actually need rather than what designers assume they should tolerate.

\subsection{Coping Strategies for Technology-Mediated Triggers (RQ2)}

Our second research question examined the coping strategies individuals with misophonia employ to manage technology-mediated trigger exposure. Participants described a range of adaptive and reactive strategies, from proactive environmental management to in-the-moment crisis responses.

Sabinson's \cite{sabinson2025blowfish} autobiographical design work demonstrates how ``designing for one's own sensory needs can reveal new possibilities for ethical behavior change in HCI.'' Her Blowfish Band ``offered a way to reclaim agency over sensory needs, often suppressed to conform to social norms.'' This insight---that sensory needs are often suppressed rather than accommodated---resonates strongly with our participants' experiences of hiding their misophonia, enduring triggers silently, and avoiding social situations rather than requesting accommodations they fear will be seen as unreasonable.

Our findings highlight the social complexity of managing triggers in different contexts. In work settings, one might invoke professional norms or ADA accommodations. Among friends and family, the dynamic is different---requests to modify behavior feel like impositions on intimacy. Participants described feeling like ``Debbie Downer'' when advocating for their needs, echoing what Baillargeon et al. \cite{baillargeon2025social} describe as the invisible labor that neurodivergent individuals perform to navigate technology-mediated spaces---``requesting essential accommodations, working remotely due to access barriers, and expressing themselves authentically in non-traditional ways.'' Platform features that make sensory needs visible and negotiable without requiring vulnerable disclosure could address this tension, transforming accessibility from an individual burden to a collective affordance.

A significant finding was the risk associated with disclosure as a coping strategy. While some participants found supportive responses when disclosing their misophonia, others experienced deliberate antagonism where disclosure led to intentional triggering. Ashley's experience exemplifies what we term \textit{epistemic trauma}---the harm that occurs when an individual's knowledge claims about their own condition are systematically dismissed, overridden, or punished by those positioned as authoritative. Her expulsion from an online community for insisting on the psychological dimensions of her experience represents more than social exclusion; it constitutes a denial of her epistemic authority over her own lived experience. This mirrors broader patterns documented in multi-stakeholder platform environments where different groups---content creators, moderators, and users---hold competing values, interests, and knowledge claims \cite{lietalmulti23}. When medical professionals claim exclusive authority to define misophonia while dismissing patient-reported experiences like visual triggers, they reproduce the epistemic injustices that drive misophones to seek online community in the first place.

The concept of user agency extends to how we frame personalization itself. Nevsky et al.'s \cite{nevskyetal25} work with people with aphasia demonstrates how highly personalized accessibility interventions can ``afford the viewer agency and the ability to become an active participant in the viewing experience.'' They argue that the ``diverse accessibility needs of a community... are better represented when working directly with individuals that embody this diversity as a vanguard for innovation---advocating for socio-technical change and fighting against ability-based segregation justified through business logic'' \cite{nevskyetal25}. This framing---of sensory-sensitive users as vanguards rather than edge cases---reorients design from accommodation to transformation, positioning misophonic users' needs not as exceptions to be tolerated but as insights that can improve platforms for everyone.

\subsection{Design Principles for Sensory-Accessible Platforms (RQ3)}

Our third research question asked what design principles can guide the development of sensory-accessible platforms that accommodate misophonic users without requiring full withdrawal from digital participation. This pattern of epistemic harm points toward the need to extend trauma-informed design principles to encompass audiovisual dimensions of platform interaction. Trauma-informed design typically focuses on content moderation, trigger warnings for distressing material, and supportive community guidelines. Our findings suggest that the \textit{sensory properties} of platforms themselves---auto-playing audio, algorithmic surfacing of ASMR content, normalized eating on camera---can constitute sources of harm for sensory-sensitive users. Moreover, platforms that fail to recognize diverse sensory experiences as legitimate may inadvertently reproduce the very epistemic dismissals that traumatize users in clinical and social contexts. Extending trauma-informed design to include audiovisual accessibility means recognizing that how content is delivered matters as much as what content is moderated, and that platform affordances encode assumptions about whose sensory experiences count as ``normal.''

Reflecting on Race et al.'s \cite{race_et_al_21} work, we argue that users should have more control over the audiovisual affordances in any technologies they use. This includes properties outlined in \cite{race_et_al_21} (e.g., duration of sounds, wavelength speed, intensity, using recognizable noises), but also extends to include setting preferences for genres like ASMR, or subtle controls over pitch, and allowing users to mute videos as a default. Our findings align with Jiang et al.'s \cite{jiang2025shifting} ADHD-inclusive design principles, which emphasize that ``designs should adapt to ADHDers' levels of focus,'' ``utilize multisensory elements to support engagement, while offering flexibility to adjust to each individual's threshold of sensory overstimulation and understimulation,'' and critically, ``involve people with ADHD throughout the design process.'' These principles---adaptation to variable needs, multisensory flexibility, and participatory involvement---apply directly to designing for misophonia.

Providing users with mechanisms to control the audio channels of individual members on a Zoom call to reduce the audio, mute, or disable the visual of any person who might be triggering them with their behavior represents a crucial step toward sensory accessibility. The provision of these controls is in line with Im et al.'s \cite{Im_um_2021} affirmative consent design principle---ingraining affirmative consent in the ways that users consume audiovisual content, much as they suggest the same for algorithmic content. Just as Lukava et al. \cite{lukava2022two} advocate for research environments that accommodate anxiety and sensory sensitivities by addressing environmental factors such as lighting intensity, noise levels, and spatial crowding, platform designers must similarly embed these considerations into everyday technology use.

Table \ref{tab:framework} summarizes our design framework, mapping principles to strategies and platforms.

% Ensure you have \usepackage{booktabs} in your preamble

\begin{table*}[h]
\centering
\caption{Design Framework for Sensory-Accessible Platforms.}
\label{tab:framework}
\small
% ACM style removes vertical bars (|)
\begin{tabular}{p{3cm}p{4cm}p{3.5cm}p{2.5cm}}
\toprule
\textbf{Principle} & \textbf{Design Strategy} & \textbf{Example Implementation} & \textbf{Primary Platforms} \\
\midrule

% Group 1
\textbf{Disaggregate Audiovisual Control} 
& Per-participant volume/video controls & Individual sliders for each call participant & Zoom, Discord, Teams \\
& Content-level audio filtering & "Reduce ASMR in my feed" toggle & TikTok, Instagram, YouTube \\
\midrule

% Group 2
\textbf{Enable Real-Time Filtering} 
& Trained audio classifiers & Detect and reduce eating sounds while preserving speech & Zoom, Discord \\
& Visual trigger detection/Per-participant video blur & Blur repetitive hand movements (misokinesia) & Zoom, Discord, social media platforms \\
\midrule

% Group 3
\textbf{Collaborative Boundary Negotiation} 
& Shared sensory profiles & Team-level sensory preferences visible to organizers & Zoom, Slack, Teams \\
& Anonymous sensory requests & Flag concerns to hosts without disclosure & Zoom, Google Meet \\
\midrule

% Group 4
\textbf{Sensory Predictability} 
& Content sensory tagging & Creators tag audio/visual properties & TikTok, Instagram, YouTube \\
& Sensory content warnings & Warning before potentially distressing content & All platforms \\
& User-controlled defaults & Set audio/video preferences that persist & All platforms \\
\midrule

\textbf{Visual Trigger Management} 
& Per-participant video blur & Selective blurring of individual video feeds & Zoom, Discord, Teams \\
& Motion detection filtering & Detect and blur repetitive movements & Zoom, social platforms \\

\bottomrule
\end{tabular}
\end{table*}

\subsubsection{Design Recommendation: Fine-Grained Audio-Visual Filtering for Inclusive Remote Interaction}
\label{sec:design-filtering}
Participants consistently described frustration with the binary nature of current audiovisual controls---either fully muted or fully exposed to potential triggers. This all-or-nothing approach forces users to choose between meaningful participation and sensory safety. The following principle addresses this gap by proposing granular, per-participant controls that preserve connection while enabling individualized sensory management.
We propose the introduction of channel-specific audio-visual control in apps like Discord, Zoom, and other real-time communication platforms. One way to implement this is by extending personal moderation tools \cite{jhaver2023users,Jhaver_et_al_23} —already familiar in social media contexts for filtering harmful or unwanted content—so that users can selectively manage sound and visual input during live calls. Such fine-grained tools have been suggested to provide users with more control over new apps to safeguard their privacy \cite{malviyaetal23}. Similarly, a misophonic user could lower or mute the volume of a specific participant or temporarily hide their video feed without affecting the overall session. Much like how personal moderation features on social platforms let individuals filter out negative comments or distressing imagery, these real-time controls would let users create individualized sensory boundaries. This approach would allow participants like Amelia to proactively avoid triggers such as chewing sounds or repetitive tapping, thereby reducing distress, sustaining participation, and preventing the need to abandon meetings or social events altogether.

\paragraph{Design Principle I: Disaggregate Audiovisual Control}

Our findings show that future designs should move from coarse-grained, platform-level controls (mute all vs. hear all) to fine-grained, individual-level controls that let users customize sensory input per participant or content source.

\textbf{Rationale:} Sensory experiences are inherently personal. What's distressing for one user may be unremarkable for another. Platform-level controls force binary choices; individual-level controls enable nuanced participation.

\textbf{Design strategies:}

\begin{itemize}
\item \textbf{Per-participant volume mixing:} Individual volume sliders for each person in a video call (Discord offers this; Zoom should follow)
\item \textbf{Selective video display:} Ability to blur, minimize, or hide specific participants' video feeds while maintaining presence indicators
\item \textbf{Audio channel separation:} Different controls for speech vs. ambient sounds from each participant
\item \textbf{Content-level filtering:} In social media, ability to set audio/visual preferences for content types (e.g., "reduce ASMR content in my feed")
\end{itemize}

\textbf{Implementation example - Zoom:} Add an "Individual Controls" panel that appears on hover over each participant's video thumbnail, with sliders for:
\begin{itemize}
\item Overall volume (0-100\%)
\item Visual clarity (clear, slight blur, heavy blur, audio-only)
\item Notification priority (alert me when this person speaks, standard, reduced)
\end{itemize}

These settings would be private to each user and persist across meetings with the same participants.

\textbf{Participant validation:} Multiple participants explicitly suggested per-person volume controls. Rhys described existing Discord controls as helpful but limited, pointing toward needs for more sophisticated individual management.

\begin{figure}[H]
    \centering
    \includegraphics[width=1\linewidth]{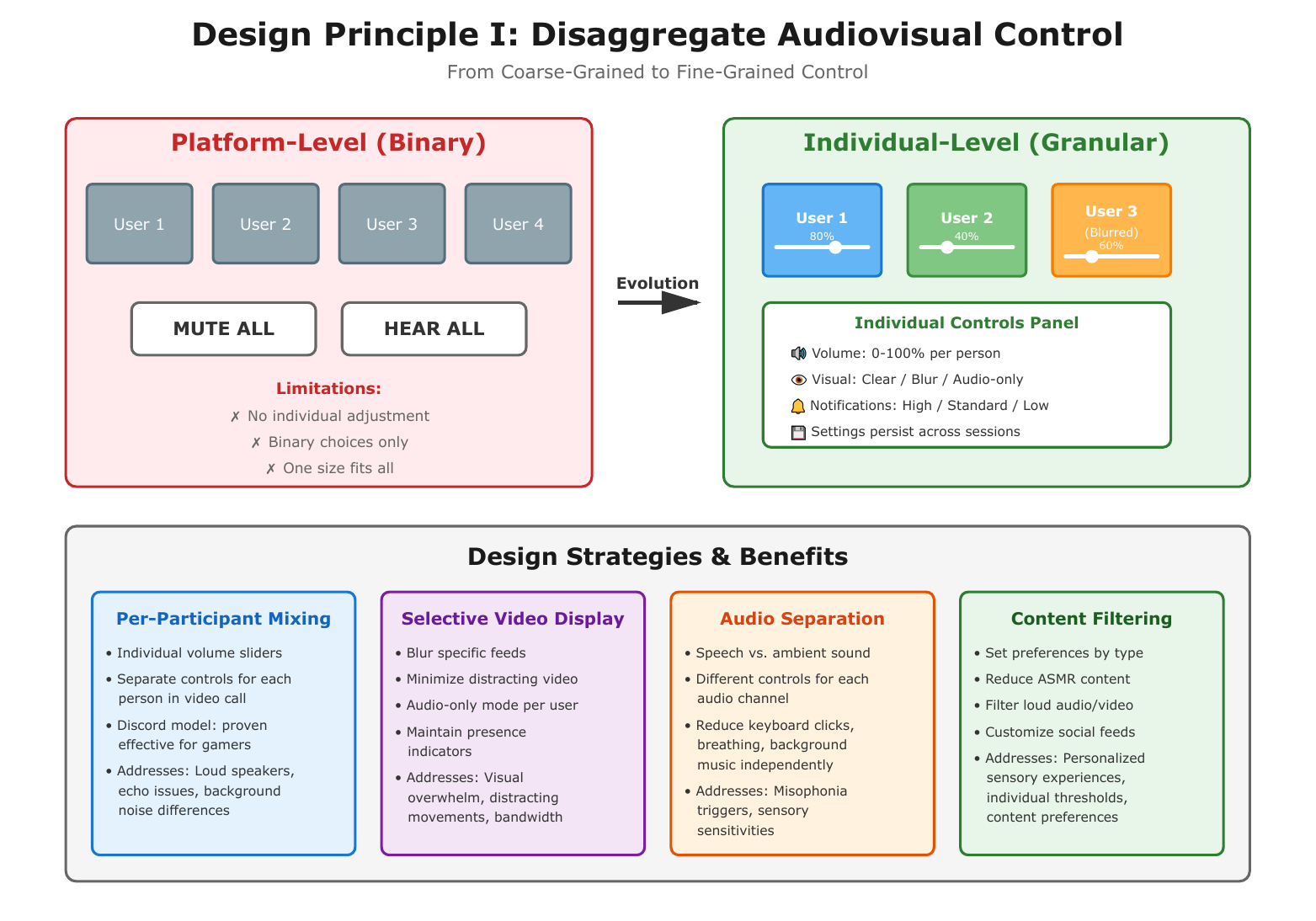}
    \caption{Design Principle I: Disaggregate Audiovisual Control. Move from coarse-grained, platform-level controls (mute all vs. hear all) to fine-grained, individual-level controls that let users customize sensory input per participant or content source. Sensory experiences are inherently personal—what's distressing for one user may be unremarkable for another. Platform-level controls force binary choices; individual-level controls enable nuanced participation through per-person volume sliders, selective video display, audio channel separation, and content-level filtering.}
    \label{fig:DIS1}
\end{figure}

\subsubsection{Design Recommendation: Real-time Automatic Detection of Triggering Sounds for Filtration}
\label{sec:design-realtime}
While per-participant controls (described above) address the granularity problem, they still require users to react to triggers after exposure. The following principle proposes proactive sensory management through machine learning-based detection and automatic filtering of trigger sounds, shifting the burden from reactive user intervention to anticipatory system support.
Different misophones have personal triggers that are different from others. Similarly, misophones can annotate and train different voices that trigger them using tools like Kim and Pardo \cite{kim_pardo_2018} focusing specifically on sounds like "coughing, laughing, screaming, sneezing, snoring, sniffling" which were predicted in earlier work with high accuracy \cite{pandey2023nocturnal}. These learned trigger models can then be used to detect and filter voices in real-time. These filters can then be used by misophones on different social media sites. A similar idea was presented by Kim et al. \cite{kim2017real} when they showed how sound data can be used to detect the presence of unmanned autonomous vehicles (UAVs) in real-time. A similar method will be used to detect specific sound snippets that represent misophonia triggers. Real-time noise filtration can then be used to filter out triggers \cite{raffaseder2008interrellation}. Given that misophones can have visual triggers, at times related to the sources of the noise (e.g., the visual of clicking a pen), we can filter out audio-visual components in real-time \cite{hou2021rule}. This is discussed in more detail in \S\ref{sec:design-visual}.

\paragraph{Design Principle II: Enable Real-Time Sensory Filtering}

Beyond volume control, platforms should filter or reduce specific sound/visual patterns that commonly trigger sensory distress (eating sounds, repetitive movements, ASMR elements) while preserving meaningful content.

\textbf{Rationale:} Binary muting removes all audio, including speech. Future designs should focus on selective filtering or reduction of triggering elements while maintaining communication. This requires treating audio as multiple simultaneous streams (speech + ambient + trigger sounds) rather than monolithic input.

\textbf{Design strategies:}

\begin{itemize}
\item \textbf{Trained audio classifiers:} Machine learning models that detect specific sound types (eating, tapping, sniffling, keyboard typing) based on user-provided training examples
\item \textbf{Real-time audio filtering:} Once detected, reduce or remove these sounds from the audio stream while preserving speech
\item \textbf{Visual trigger detection:} Identify and optionally blur visual triggers like repetitive hand movements (misokinesia \cite{jaswal2021misokinesia})
\item \textbf{User-trainable models:} Let users annotate their personal triggers, training personalized filters
\end{itemize}

\textbf{Implementation example - Discord/Zoom:} "Sensory Filter" feature with two modes:

\textit{Preset mode:} Toggle common trigger categories:
\begin{itemize}
\item Reduce eating sounds: ON
\item Reduce keyboard/typing: OFF  
\item Reduce repetitive tapping: ON
\item Reduce sniffling/throat clearing: ON
\item \textit{Custom mode:} Record 3-5 second samples of specific triggers ("my roommate's pen clicking," "this colleague's gum chewing"). System learns these patterns and filters them in real-time.
\end{itemize}

\textbf{Technical feasibility:} Audio event detection research \cite{kim_pardo_2018, pandey2023nocturnal} demonstrates high accuracy (>85\%) for classifying sounds like coughing, sniffling, and eating. Real-time filtering techniques \cite{raffaseder2008interrellation, hou2021rule} can reduce specific audio components without degrading speech. These technologies exist; they need integration into user-facing features.

\begin{figure}[H]
    \centering
    \includegraphics[width=1\linewidth]{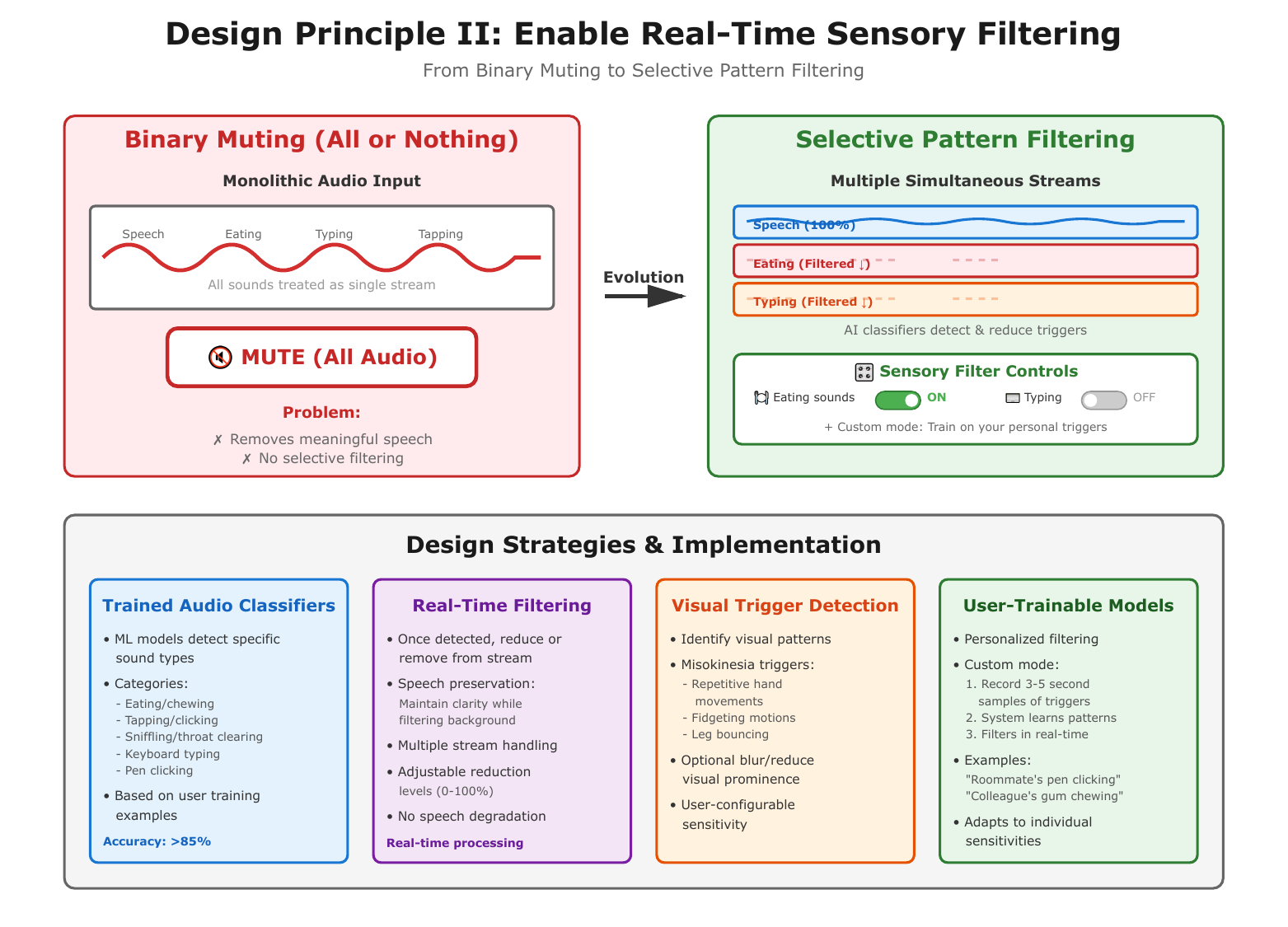}
    \caption{Design Principle II: Enable Real-Time Sensory Filtering. Beyond volume control, platforms should filter or reduce specific sound and visual patterns that commonly trigger sensory distress (eating sounds, repetitive movements, ASMR elements) while preserving meaningful content. Binary muting removes all audio, including speech—what's needed is selective reduction of triggering elements while maintaining communication. This requires treating audio as multiple simultaneous streams (speech + ambient + trigger sounds) rather than monolithic input, using trained audio classifiers, real-time filtering techniques, visual trigger detection, and user-trainable models to achieve >85\% accuracy in identifying and reducing specific triggers without degrading speech quality.}
    \label{fig:placeholder}
\end{figure}

\subsubsection{Design Recommendation: Sensory Predictability and Content Warnings}
\label{sec:design-predictability}
Unpredictability amplifies distress---participants described heightened anxiety not only from triggers themselves but from uncertainty about when triggers might occur. The following principle addresses this temporal dimension of sensory accessibility, proposing content tagging, warnings, and preview mechanisms that allow users to anticipate and prepare for potential sensory challenges.

\paragraph{Principle III: Design for Sensory Predictability}

Uncertainty increases distress and in some causes burnout that pushes many users out of online platforms altogether -- this is especially significant in uncertain health contexts \cite{joyammari25}. Platforms should make sensory experiences predictable and controllable through content warnings, sensory tags, and user-controlled defaults.

\textbf{Rationale:} Participants described distress from unexpected triggers—algorithms surfacing ASMR content, colleagues suddenly eating on camera. Predictability reduces anxiety and enables proactive management.

\textbf{Design strategies:}

\begin{itemize}
\item \textbf{Content sensory tagging:} Creators tag content with sensory properties (loud audio, eating visuals, flashing lights)
\item \textbf{Sensory content warnings:} Similar to content warnings for violence, warn users about potentially distressing sensory content
\item \textbf{User-controlled defaults:} Let users set sensory defaults (e.g., "always mute ASMR content," "auto-blur eating videos," "default to camera off in meetings")
\item \textbf{Preview before full exposure:} Allow users to preview audio levels and visual content before committing to watch/participate
\end{itemize}

\textbf{Implementation example - TikTok:} Add sensory tags to content creation:
\begin{itemize}
\item Audio: [Eating sounds] [ASMR] [Loud noises] [Repetitive sounds] [Whispering]
\item Visual: [Eating visuals] [Repetitive motion] [Close-up faces] [Mouth sounds]
\end{itemize}

Users set preferences: "Minimize [Eating sounds] and [ASMR]." Algorithm adjusts accordingly, similar to "Not Interested" feedback but specifically for sensory properties.

Content with sensory tags shows warnings: "This video contains eating sounds. [Watch anyway] [Skip]"

\begin{figure}[h]
    \centering
    \includegraphics[width=1\linewidth]{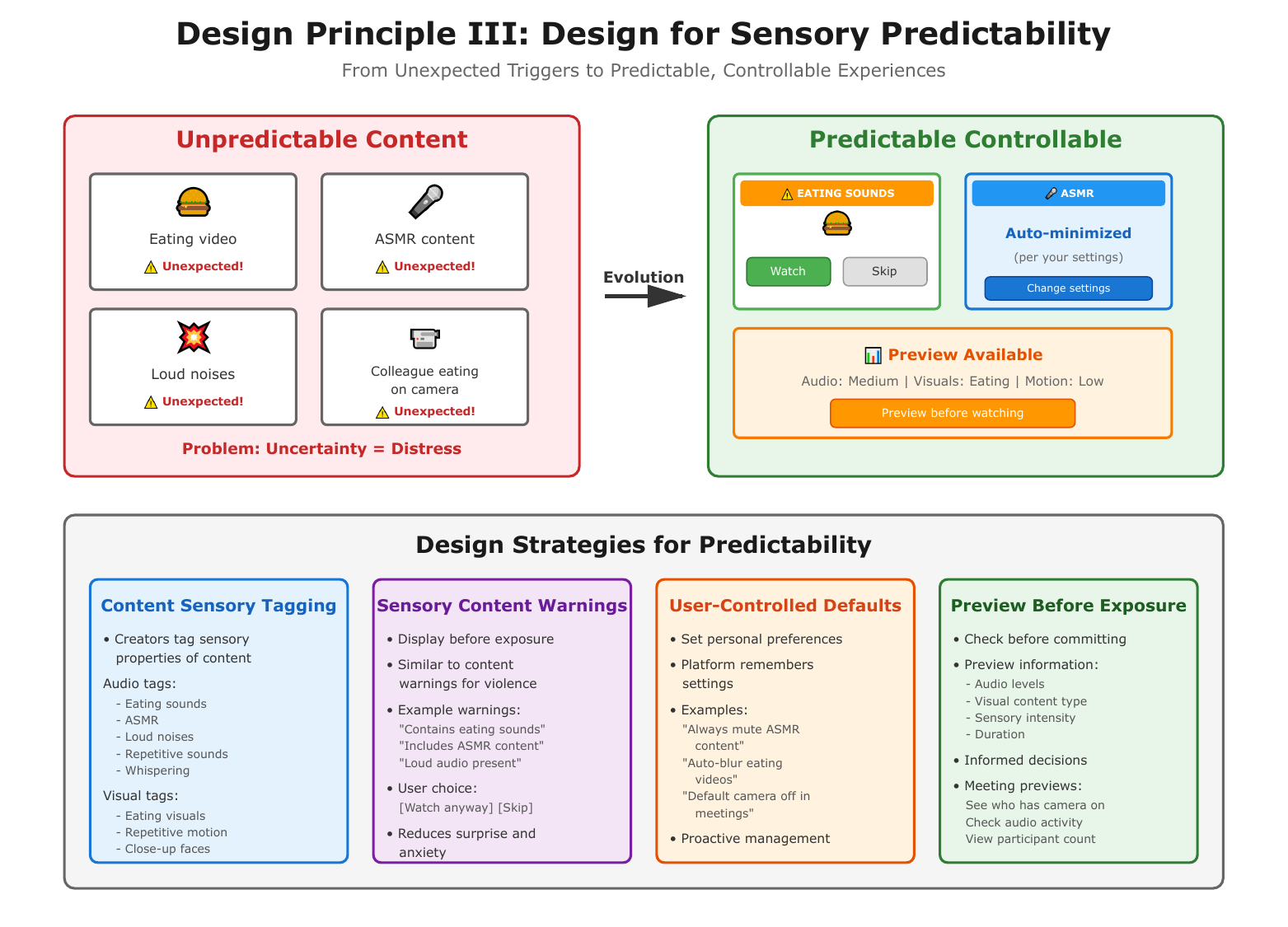}
    \caption{Design Principle III: Design for Sensory Predictability.
Uncertainty increases distress, especially in healthcare contexts. Platforms should make sensory experiences predictable and controllable through content warnings, sensory tags, and user-controlled defaults. Participants described distress from unexpected triggers—algorithms surfacing ASMR content, colleagues suddenly eating on camera. Predictability reduces anxiety and enables proactive management through content sensory tagging (creators label audio and visual properties), sensory content warnings (similar to violence warnings, alerting users before exposure), user-controlled defaults (set preferences like "always mute ASMR content" or "auto-blur eating videos"), and preview options that allow users to check audio levels and visual content before committing to watch or participate.}
    \label{fig:DIS3}
\end{figure}

\subsubsection{Design Recommendation: Joint Account Audio-Visual Preference Tools}
\label{sec:design-collaborative}
Sensory accessibility extends beyond individual user-platform interactions to encompass shared social contexts. Participants described the challenge of advocating for their needs in group settings without appearing demanding or unreasonable. The following principle addresses this social dimension, proposing collaborative preference-setting mechanisms that distribute the burden of accommodation across groups rather than placing it solely on sensory-sensitive individuals.
Misophones have to negotiate daily routines with co-workers, family, roommates, and friends  if they are sources of misophonia triggers. In other words, misophones might have their audiovisual boundaries perturbed through their interactions, whether technologically mediated or not, with others in their social groups. We conceive of this concept building on the boundary-turbulence framework introduced by Petronio \cite{petronio2002boundaries}. Petronio's framework focuses on privacy boundaries that might be perturbed because of the varying privacy needs between members of a social group. In this case, there are specific audiovisual needs for misophones that are usually not shared by their social peers. For example, listening to TikTok videos within earshot may perturb misophones' audiovisual boundaries. Eating dinner while on yet another Zoom meeting might be acceptable to most people but perturb the audiovisual boundaries of misophones on the same call. Creating joint account preference tools in which designers can allow for the creation of ``user cirlces'' where groups can setup shared audio-visual preferences could reduce frictions between misophones and others in their social circles thus reducing instances of technology non-use.

\paragraph{Principle IV: Support Collaborative Boundary Negotiation}

\textbf{Principle:} Sensory accessibility isn't just about individual users managing their experience—it requires negotiating boundaries with others who share the space. Platforms should facilitate these negotiations rather than forcing users to choose between self-advocacy or silence.

\textbf{Rationale:} Participants described difficulty advocating for their needs, often feeling like "Debbie Downer" (Amelia) or worrying about seeming demanding. Platform features can make sensory needs visible and negotiable without requiring vulnerable disclosure.

\textbf{Design strategies:}

\begin{itemize}
\item \textbf{Shared sensory profiles:} Groups (work teams, friend circles, families) can create shared profiles noting common sensory sensitivities
\item \textbf{Pre-meeting sensory preferences:} Before meetings, participants can see aggregated sensory preferences (e.g., "2 participants prefer camera-optional, 1 requests minimal background noise")
\item \textbf{Gentle nudges:} Platform suggests considerate behaviors based on group preferences (e.g., "This meeting includes participants sensitive to eating sounds—consider eating before/after")
\item \textbf{Anonymous sensory requests:} Users can flag sensory concerns anonymously to meeting hosts, who can address them generally without singling out individuals
\end{itemize}

\textbf{Implementation example - Zoom "Team Sensory Profile":} 

Organizations or social groups create a team profile where members optionally indicate:
\begin{itemize}
\item Audio sensitivities (eating, typing, background noise)
\item Visual sensitivities (bright lights, motion, patterns)
\item Preference for camera use (always on, optional, prefer off)
\item Preferred meeting formats (video, audio-only, chat-based)
\end{itemize}

When scheduling meetings, the organizer sees an anonymized summary: "Your team prefers: camera-optional (4), minimal background noise (3), no eating on camera (2)." Meeting invitations can include these preferences: "This team values camera-optional participation and minimizing eating sounds during calls."

\textbf{Ethical consideration:} This approach requires careful privacy design. Users must control what information is shared and with whom. Anonymized aggregation helps, but platforms must avoid creating pressure to disclose sensory needs or stigmatizing those who do.

\begin{figure}[H]
    \centering
    \includegraphics[width=1\linewidth]{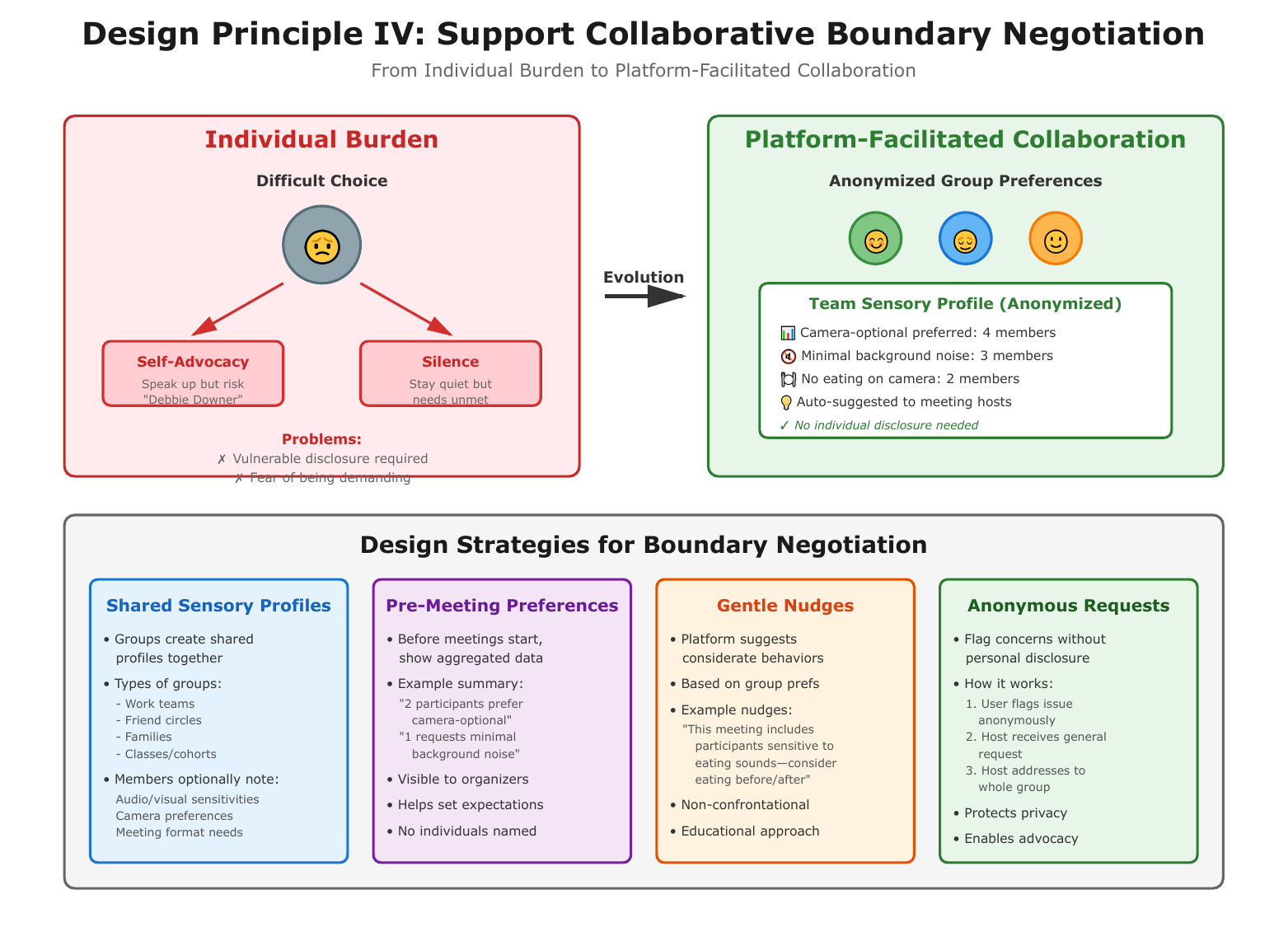}
    \caption{Design Principle IV: Support Collaborative Boundary Negotiation. Sensory accessibility is not just about individual users managing their experience—it requires negotiating boundaries with others who share the space. Platforms should facilitate these negotiations rather than forcing users to choose between self-advocacy or silence. Participants described difficulty advocating for their needs, often feeling like "Debbie Downer" or worrying about seeming demanding. Platform features can make sensory needs visible and negotiable without requiring vulnerable disclosure through shared sensory profiles (work teams, friend circles, families noting common sensitivities), pre-meeting sensory preferences (anonymized summaries like "2 participants prefer camera-optional"), gentle nudges (platform suggests considerate behaviors based on group preferences), and anonymous sensory requests (users flag concerns to hosts who address them generally without singling out individuals). This approach requires careful privacy design where users control what information is shared and with whom.}
    \label{fig:DIS4}
\end{figure}

\subsubsection{Design Recommendation: Selective Video Management for Visual Trigger Mitigation}
\label{sec:design-visual}
Misokinesia---sensitivity to visual stimuli such as repetitive movements, eating visuals, or fidgeting---frequently co-occurs with misophonia, yet platform designs overwhelmingly focus on auditory accessibility while neglecting visual dimensions. Participants in our study described visual triggers as distinct from but often accompanying auditory ones, indicating that comprehensive sensory accessibility must address both modalities. This section proposes design strategies for granular visual control that preserve social presence while reducing sensory distress.

\paragraph{Principle V: Address Visual Triggers Through Selective Video Management}

\textbf{Principle:} Misophonia is often accompanied by misokinesia---sensitivity to visual triggers such as repetitive movements, eating visuals, or fidgeting behaviors. Platforms should provide granular control over visual input, allowing users to selectively blur, minimize, or filter video feeds without fully disengaging from visual communication.

\textbf{Rationale:} Participants described visual triggers as distinct from but often co-occurring with auditory ones. Nathan noted that seeing someone eat on Zoom triggered distress even when audio was muted, rating visual triggers as ``a two or three'' compared to auditory triggers at ``a nine.'' Kristie described being triggered by repetitive movements like neck rubbing or fidgeting, expressing a wish for ``blinders'' to avoid seeing triggering behaviors. Amelia noted her husband's knee bouncing as a persistent visual trigger. These experiences indicate that audio-only solutions address only part of the sensory accessibility challenge.

\textbf{Design strategies:}

\begin{itemize}
\item \textbf{Per-participant video blur:} Allow users to apply selective blur to individual video feeds, reducing visual detail while maintaining awareness of presence and general activity
\item \textbf{Motion detection filtering:} Implement computer vision to detect and selectively blur repetitive movements (hand fidgeting, leg bouncing, hair twirling) while preserving facial expressions and gestures essential for communication
\item \textbf{Eating detection and response:} Automatically detect eating behaviors and offer users the option to temporarily blur that participant's feed or receive a warning before the video loads
\item \textbf{User-defined visual triggers:} Allow users to specify visual patterns that trigger distress, with the platform learning to identify and filter similar content
\item \textbf{Thumbnail-only mode:} Option to view participants as static thumbnails rather than live video, with indicators showing when someone is speaking
\end{itemize}

\textbf{Implementation example - Zoom ``Visual Comfort Controls'':}

Add a ``Visual Comfort'' panel accessible from each participant's video tile:
\begin{itemize}
\item \textbf{Blur level:} None / Light (softens details) / Medium (obscures features) / Heavy (silhouette only)
\item \textbf{Motion filter:} Off / Reduce peripheral motion / Reduce all motion
\item \textbf{Eating alert:} Off / Notify me / Auto-blur during eating
\item \textbf{View mode:} Live video / Refreshing snapshot (updates every 5 seconds) / Static avatar
\end{itemize}

These settings remain private to each user and persist across sessions with the same participants. Users can set defaults (e.g., ``always apply light blur to all participants'') while retaining the ability to adjust individual feeds.

\textbf{Technical feasibility:} Computer vision research has demonstrated reliable detection of eating behaviors, hand movements, and repetitive motions. Real-time video processing for selective blurring is computationally feasible on modern devices. Platforms already implement background blur and virtual backgrounds, indicating the technical infrastructure exists for more sophisticated per-region video processing.

\textbf{Ethical consideration:} Visual filtering raises questions about social presence and authentic interaction. Platforms should frame these features as accessibility tools rather than ways to ``hide'' from colleagues. Clear communication about the purpose---reducing sensory distress while maintaining connection---can help normalize their use without stigma.

\begin{figure}[H]
    \centering
    \includegraphics[width=1\linewidth]{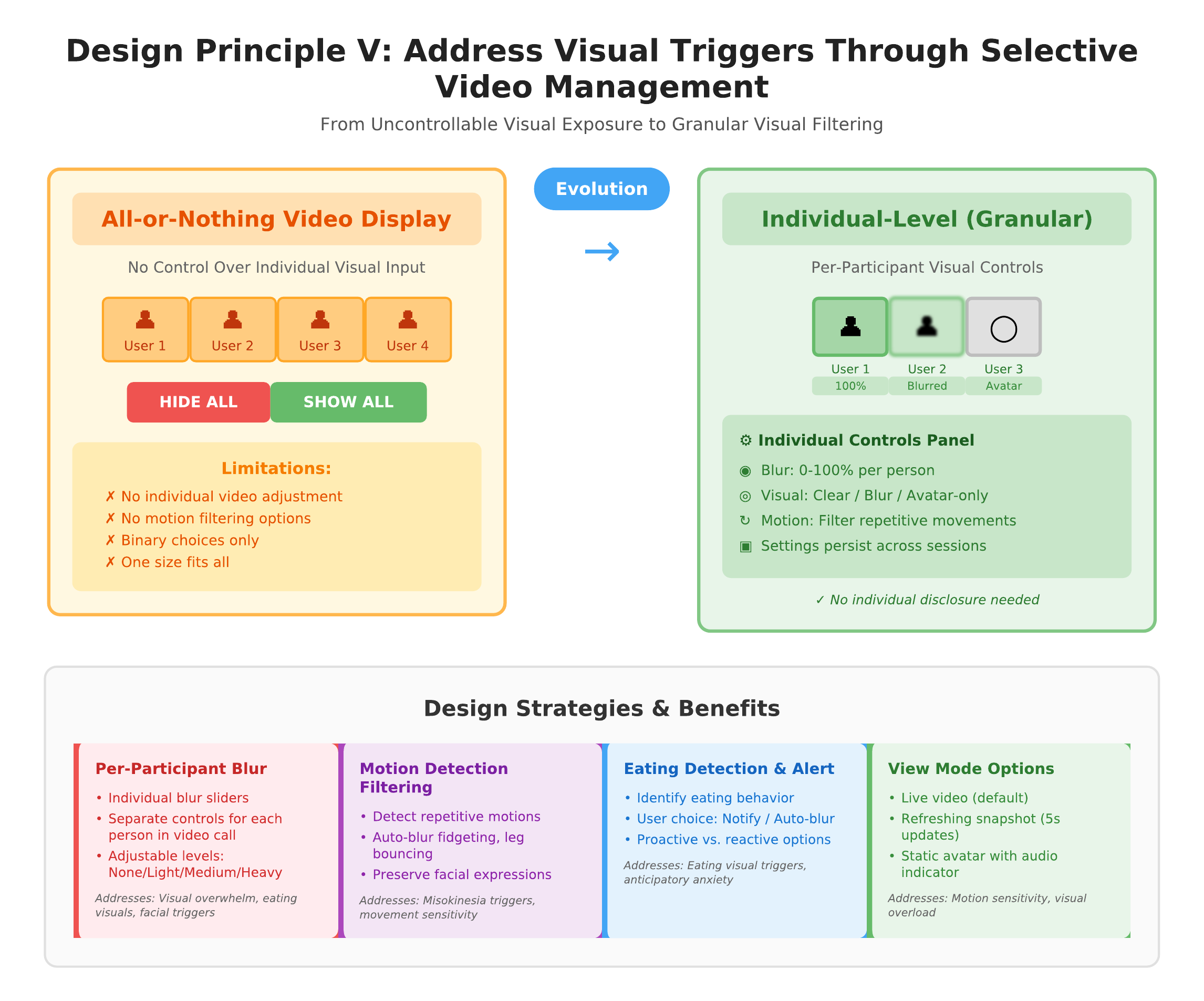}
    \caption{Design Principle V: Address Visual Triggers Through Selective Video Management. Misophonia often co-occurs with misokinesia---sensitivity to visual triggers such as repetitive movements, eating visuals, or fidgeting. Participants described visual triggers as distinct from but often accompanying auditory ones, with one noting that seeing someone eat triggered distress even when audio was muted. Platforms should provide granular control over visual input through per-participant video blur (selective blur levels while maintaining presence awareness), motion detection filtering (computer vision to detect and blur repetitive movements while preserving communication-essential gestures), eating detection (automatic detection with options to blur or warn), user-defined visual triggers (learning to identify and filter specified patterns), and thumbnail-only mode (static images with speaking indicators). These features leverage existing video processing infrastructure while framing visual filtering as an accessibility tool to reduce sensory distress while maintaining connection.}
    \label{fig:DIS5}
\end{figure}
\section{Limitations and Future Work}
This study is based on in-depth qualitative interviews with a relatively small group of U.S.-based participants recruited from online communities. While these rich narratives illuminate how misophonia shapes technology use and avoidance, the findings are not statistically generalizable. Future research should complement this work with large-scale surveys to capture the prevalence and diversity of technology non-use among people with misophonia. Such surveys could reveal broader patterns across demographics, geographic regions, and technology ecosystems, and help prioritize features for design interventions.

Beyond expanding the sample, participatory design with misophonic users offers an important next step. Engaging participants directly in co-design workshops or iterative prototyping would allow researchers and designers to collaboratively refine tools such as channel-specific audio-visual controls, real-time trigger detection, and shared preference settings. These approaches would ensure that future technologies not only reduce distress but also reflect the lived expertise and agency of the communities they aim to support.

\section{Conclusion}
This study shows how misophonia influences technology use and avoidance, particularly on social and communication platforms whose audiovisual affordances intensify sensory triggers. By surfacing these challenges, we extend prior research on neurodivergence and mediated interaction. We recommend design features that give users finer control over sound and visual inputs, enable real-time filtering of personal triggers, and support collaborative boundary setting. Such measures can reduce distress and broaden participation for people with misophonia, fostering more inclusive digital environments across work, leisure, and social life.

\bibliographystyle{ACM-Reference-Format}
\bibliography{sample-base}


\begin{thebibliography}{82}

%%% the Web address.

\ifx \showCODEN    \undefined \def \showCODEN     #1{\unskip}     \fi
\ifx \showISBNx    \undefined \def \showISBNx     #1{\unskip}     \fi
\ifx \showISBNxiii \undefined \def \showISBNxiii  #1{\unskip}     \fi
\ifx \showISSN     \undefined \def \showISSN      #1{\unskip}     \fi
\ifx \showLCCN     \undefined \def \showLCCN      #1{\unskip}     \fi
\ifx \shownote     \undefined \def \shownote      #1{#1}          \fi
\ifx \showarticletitle \undefined \def \showarticletitle #1{#1}   \fi
\ifx \showURL      \undefined \def \showURL       {\relax}        \fi
% invisible to TeX
\providecommand\bibfield[2]{#2}
\providecommand\bibinfo[2]{#2}
\providecommand\natexlab[1]{#1}
\providecommand\showeprint[2][]{arXiv:#2}

\bibitem[Baillargeon et~al\mbox{.}(2025)]%
        {baillargeon2025social}
\bibfield{author}{\bibinfo{person}{Philip Baillargeon}, \bibinfo{person}{Jina Yoon}, {and} \bibinfo{person}{Amy Zhang}.} \bibinfo{year}{2025}\natexlab{}.
\newblock \showarticletitle{Who Puts the "Social" in "Social Computing"?: Using A Neurodiversity Framing to Review Social Computing Research}.
\newblock \bibinfo{journal}{\emph{Proc. ACM Hum.-Comput. Interact.}} \bibinfo{volume}{9}, \bibinfo{number}{2}, Article \bibinfo{articleno}{CSCW208} (\bibinfo{date}{May} \bibinfo{year}{2025}), \bibinfo{numpages}{44}~pages.
\newblock
\href{https://doi.org/10.1145/3711106}{doi:\nolinkurl{10.1145/3711106}}


\bibitem[Balka and Wagner(2006)]%
        {balka2006making}
\bibfield{author}{\bibinfo{person}{Ellen Balka} {and} \bibinfo{person}{Ina Wagner}.} \bibinfo{year}{2006}\natexlab{}.
\newblock \showarticletitle{Making things work: dimensions of configurability as appropriation work}. In \bibinfo{booktitle}{\emph{Proceedings of the 2006 20th anniversary conference on Computer supported cooperative work}}. \bibinfo{pages}{229--238}.
\newblock


\bibitem[Barocas et~al\mbox{.}(2020)]%
        {barocas2020}
\bibfield{author}{\bibinfo{person}{Solon Barocas}, \bibinfo{person}{Asia~J. Biega}, \bibinfo{person}{Benjamin Fish}, \bibinfo{person}{Jundefineddrzej Niklas}, {and} \bibinfo{person}{Luke Stark}.} \bibinfo{year}{2020}\natexlab{}.
\newblock \showarticletitle{When not to design, build, or deploy}. In \bibinfo{booktitle}{\emph{Proceedings of the 2020 Conference on Fairness, Accountability, and Transparency}} (Barcelona, Spain) \emph{(\bibinfo{series}{FAT* '20})}. \bibinfo{publisher}{Association for Computing Machinery}, \bibinfo{address}{New York, NY, USA}, \bibinfo{pages}{695}.
\newblock
\showISBNx{9781450369367}
\href{https://doi.org/10.1145/3351095.3375691}{doi:\nolinkurl{10.1145/3351095.3375691}}


\bibitem[Baumer et~al\mbox{.}(2013)]%
        {baumer_et_al_13}
\bibfield{author}{\bibinfo{person}{Eric~P.S. Baumer}, \bibinfo{person}{Phil Adams}, \bibinfo{person}{Vera~D. Khovanskaya}, \bibinfo{person}{Tony~C. Liao}, \bibinfo{person}{Madeline~E. Smith}, \bibinfo{person}{Victoria Schwanda~Sosik}, {and} \bibinfo{person}{Kaiton Williams}.} \bibinfo{year}{2013}\natexlab{}.
\newblock \showarticletitle{Limiting, leaving, and (re)lapsing: an exploration of facebook non-use practices and experiences}. In \bibinfo{booktitle}{\emph{Proceedings of the SIGCHI Conference on Human Factors in Computing Systems}} (Paris, France) \emph{(\bibinfo{series}{CHI '13})}. \bibinfo{publisher}{Association for Computing Machinery}, \bibinfo{address}{New York, NY, USA}, \bibinfo{pages}{3257–3266}.
\newblock
\showISBNx{9781450318990}
\href{https://doi.org/10.1145/2470654.2466446}{doi:\nolinkurl{10.1145/2470654.2466446}}


\bibitem[Baumer et~al\mbox{.}(2014)]%
        {baumer_et_al_14}
\bibfield{author}{\bibinfo{person}{Eric~P.S. Baumer}, \bibinfo{person}{Morgan~G. Ames}, \bibinfo{person}{Jed~R. Brubaker}, \bibinfo{person}{Jenna Burrell}, {and} \bibinfo{person}{Paul Dourish}.} \bibinfo{year}{2014}\natexlab{}.
\newblock \showarticletitle{Refusing, limiting, departing: why we should study technology non-use}. In \bibinfo{booktitle}{\emph{CHI '14 Extended Abstracts on Human Factors in Computing Systems}} (Toronto, Ontario, Canada) \emph{(\bibinfo{series}{CHI EA '14})}. \bibinfo{publisher}{Association for Computing Machinery}, \bibinfo{address}{New York, NY, USA}, \bibinfo{pages}{65–68}.
\newblock
\showISBNx{9781450324748}
\href{https://doi.org/10.1145/2559206.2559224}{doi:\nolinkurl{10.1145/2559206.2559224}}


\bibitem[Becker et~al\mbox{.}(2017)]%
        {becker2019audiovisual}
\bibfield{author}{\bibinfo{person}{Valdecir Becker}, \bibinfo{person}{Daniel Gambaro}, \bibinfo{person}{Thais Saraiva~Ramos}, {and} \bibinfo{person}{Rafael Moura~Toscano}.} \bibinfo{year}{2017}\natexlab{}.
\newblock \showarticletitle{Audiovisual design: introducing ‘media affordances’ as a relevant concept for the development of a new communication model}. In \bibinfo{booktitle}{\emph{Iberoamerican Conference on Applications and Usability of Interactive TV}}. Springer, \bibinfo{pages}{17--31}.
\newblock


\bibitem[Bennett and Rosner(2019)]%
        {bennett2019promise}
\bibfield{author}{\bibinfo{person}{Cynthia~L. Bennett} {and} \bibinfo{person}{Daniela~K. Rosner}.} \bibinfo{year}{2019}\natexlab{}.
\newblock \showarticletitle{The Promise of Empathy: Design, Disability, and Knowing the "Other"}. In \bibinfo{booktitle}{\emph{Proceedings of the 2019 CHI Conference on Human Factors in Computing Systems}} (Glasgow, Scotland Uk) \emph{(\bibinfo{series}{CHI '19})}. \bibinfo{publisher}{Association for Computing Machinery}, \bibinfo{address}{New York, NY, USA}, \bibinfo{pages}{1–13}.
\newblock
\showISBNx{9781450359702}
\href{https://doi.org/10.1145/3290605.3300528}{doi:\nolinkurl{10.1145/3290605.3300528}}


\bibitem[Brewer et~al\mbox{.}(2021)]%
        {brewer_et_al_21}
\bibfield{author}{\bibinfo{person}{Robin~N. Brewer}, \bibinfo{person}{Sarita Schoenebeck}, \bibinfo{person}{Kerry Lee}, {and} \bibinfo{person}{Haripriya Suryadevara}.} \bibinfo{year}{2021}\natexlab{}.
\newblock \showarticletitle{Challenging Passive Social Media Use: Older Adults as Caregivers Online}.
\newblock \bibinfo{journal}{\emph{Proc. ACM Hum.-Comput. Interact.}} \bibinfo{volume}{5}, \bibinfo{number}{CSCW1}, Article \bibinfo{articleno}{123} (\bibinfo{date}{April} \bibinfo{year}{2021}), \bibinfo{numpages}{20}~pages.
\newblock
\href{https://doi.org/10.1145/3449197}{doi:\nolinkurl{10.1145/3449197}}


\bibitem[Cavanna and Seri(2015)]%
        {cavanna2015misophonia}
\bibfield{author}{\bibinfo{person}{Andrea~E Cavanna} {and} \bibinfo{person}{Stefano Seri}.} \bibinfo{year}{2015}\natexlab{}.
\newblock \showarticletitle{Misophonia: current perspectives}.
\newblock \bibinfo{journal}{\emph{Neuropsychiatric disease and treatment}} (\bibinfo{year}{2015}), \bibinfo{pages}{2117--2123}.
\newblock


\bibitem[Cha and Wong(2025)]%
        {chatetal25}
\bibfield{author}{\bibinfo{person}{Inha Cha} {and} \bibinfo{person}{Richmond~Y. Wong}.} \bibinfo{year}{2025}\natexlab{}.
\newblock \showarticletitle{Understanding Socio-technical Factors Configuring AI Non-Use in UX Work Practices}. In \bibinfo{booktitle}{\emph{Proceedings of the 2025 CHI Conference on Human Factors in Computing Systems}} \emph{(\bibinfo{series}{CHI '25})}. \bibinfo{publisher}{Association for Computing Machinery}, \bibinfo{address}{New York, NY, USA}, Article \bibinfo{articleno}{1110}, \bibinfo{numpages}{17}~pages.
\newblock
\showISBNx{9798400713941}
\href{https://doi.org/10.1145/3706598.3713140}{doi:\nolinkurl{10.1145/3706598.3713140}}


\bibitem[Chen et~al\mbox{.}(2022)]%
        {chen_trauma-informed_2022}
\bibfield{author}{\bibinfo{person}{Janet~X. Chen}, \bibinfo{person}{Allison McDonald}, \bibinfo{person}{Yixin Zou}, \bibinfo{person}{Emily Tseng}, \bibinfo{person}{Kevin~A Roundy}, \bibinfo{person}{Acar Tamersoy}, \bibinfo{person}{Florian Schaub}, \bibinfo{person}{Thomas Ristenpart}, {and} \bibinfo{person}{Nicola Dell}.} \bibinfo{year}{2022}\natexlab{}.
\newblock \showarticletitle{Trauma-Informed Computing: Towards Safer Technology Experiences for All}. In \bibinfo{booktitle}{\emph{Proceedings of the 2022 CHI Conference on Human Factors in Computing Systems}} (New Orleans, LA, USA) \emph{(\bibinfo{series}{CHI '22})}. \bibinfo{publisher}{Association for Computing Machinery}, \bibinfo{address}{New York, NY, USA}, Article \bibinfo{articleno}{544}, \bibinfo{numpages}{20}~pages.
\newblock
\showISBNx{9781450391573}
\href{https://doi.org/10.1145/3491102.3517475}{doi:\nolinkurl{10.1145/3491102.3517475}}


\bibitem[Clarke and Braun(2013)]%
        {clarke2013successful}
\bibfield{author}{\bibinfo{person}{Victoria Clarke} {and} \bibinfo{person}{Virginia Braun}.} \bibinfo{year}{2013}\natexlab{}.
\newblock \showarticletitle{Successful qualitative research: A practical guide for beginners}.
\newblock \bibinfo{journal}{\emph{Successful qualitative research}} (\bibinfo{year}{2013}), \bibinfo{pages}{1--400}.
\newblock


\bibitem[Das et~al\mbox{.}(2021)]%
        {das_et_al_21}
\bibfield{author}{\bibinfo{person}{Maitraye Das}, \bibinfo{person}{John Tang}, \bibinfo{person}{Kathryn~E. Ringland}, {and} \bibinfo{person}{Anne~Marie Piper}.} \bibinfo{year}{2021}\natexlab{}.
\newblock \showarticletitle{Towards Accessible Remote Work: Understanding Work-from-Home Practices of Neurodivergent Professionals}.
\newblock \bibinfo{journal}{\emph{Proc. ACM Hum.-Comput. Interact.}} \bibinfo{volume}{5}, \bibinfo{number}{CSCW1}, Article \bibinfo{articleno}{183} (\bibinfo{date}{apr} \bibinfo{year}{2021}), \bibinfo{numpages}{30}~pages.
\newblock
\href{https://doi.org/10.1145/3449282}{doi:\nolinkurl{10.1145/3449282}}


\bibitem[Densmore(2012)]%
        {densmore2012}
\bibfield{author}{\bibinfo{person}{Melissa Densmore}.} \bibinfo{year}{2012}\natexlab{}.
\newblock \showarticletitle{Claim mobile: when to fail a technology}. In \bibinfo{booktitle}{\emph{Proceedings of the SIGCHI Conference on Human Factors in Computing Systems}} (Austin, Texas, USA) \emph{(\bibinfo{series}{CHI '12})}. \bibinfo{publisher}{Association for Computing Machinery}, \bibinfo{address}{New York, NY, USA}, \bibinfo{pages}{1833–1842}.
\newblock
\showISBNx{9781450310154}
\href{https://doi.org/10.1145/2207676.2208319}{doi:\nolinkurl{10.1145/2207676.2208319}}


\bibitem[DiSalvo(2015)]%
        {disalvo2015adversarial}
\bibfield{author}{\bibinfo{person}{Carl DiSalvo}.} \bibinfo{year}{2015}\natexlab{}.
\newblock \bibinfo{booktitle}{\emph{Adversarial design}}.
\newblock \bibinfo{publisher}{Mit Press}.
\newblock


\bibitem[Dourish et~al\mbox{.}(2020)]%
        {dourishetal2020}
\bibfield{author}{\bibinfo{person}{Paul Dourish}, \bibinfo{person}{Christopher Lawrence}, \bibinfo{person}{Tuck~Wah Leong}, {and} \bibinfo{person}{Greg Wadley}.} \bibinfo{year}{2020}\natexlab{}.
\newblock \showarticletitle{On Being Iterated: The Affective Demands of Design Participation}. In \bibinfo{booktitle}{\emph{Proceedings of the 2020 CHI Conference on Human Factors in Computing Systems}} (Honolulu, HI, USA) \emph{(\bibinfo{series}{CHI '20})}. \bibinfo{publisher}{Association for Computing Machinery}, \bibinfo{address}{New York, NY, USA}, \bibinfo{pages}{1–11}.
\newblock
\showISBNx{9781450367080}
\href{https://doi.org/10.1145/3313831.3376545}{doi:\nolinkurl{10.1145/3313831.3376545}}


\bibitem[Eglash et~al\mbox{.}(2006)]%
        {eglash2006technology}
\bibfield{author}{\bibinfo{person}{Ron Eglash} {et~al\mbox{.}}} \bibinfo{year}{2006}\natexlab{}.
\newblock \showarticletitle{Technology as material culture}.
\newblock \bibinfo{journal}{\emph{Handbook of material culture}} (\bibinfo{year}{2006}), \bibinfo{pages}{329--340}.
\newblock


\bibitem[Epstein et~al\mbox{.}(2020)]%
        {epstein_2020}
\bibfield{author}{\bibinfo{person}{Daniel~A. Epstein}, \bibinfo{person}{Clara Caldeira}, \bibinfo{person}{Mayara~Costa Figueiredo}, \bibinfo{person}{Xi Lu}, \bibinfo{person}{Lucas~M. Silva}, \bibinfo{person}{Lucretia Williams}, \bibinfo{person}{Jong~Ho Lee}, \bibinfo{person}{Qingyang Li}, \bibinfo{person}{Simran Ahuja}, \bibinfo{person}{Qiuer Chen}, \bibinfo{person}{Payam Dowlatyari}, \bibinfo{person}{Craig Hilby}, \bibinfo{person}{Sazeda Sultana}, \bibinfo{person}{Elizabeth~V. Eikey}, {and} \bibinfo{person}{Yunan Chen}.} \bibinfo{year}{2020}\natexlab{}.
\newblock \showarticletitle{Mapping and Taking Stock of the Personal Informatics Literature}.
\newblock \bibinfo{journal}{\emph{Proc. ACM Interact. Mob. Wearable Ubiquitous Technol.}} \bibinfo{volume}{4}, \bibinfo{number}{4}, Article \bibinfo{articleno}{126} (\bibinfo{date}{Dec.} \bibinfo{year}{2020}), \bibinfo{numpages}{38}~pages.
\newblock
\href{https://doi.org/10.1145/3432231}{doi:\nolinkurl{10.1145/3432231}}


\bibitem[Frauenberger(2015)]%
        {frauenberger2015disability}
\bibfield{author}{\bibinfo{person}{Christopher Frauenberger}.} \bibinfo{year}{2015}\natexlab{}.
\newblock \showarticletitle{Disability and Technology: A Critical Realist Perspective}. In \bibinfo{booktitle}{\emph{Proceedings of the 17th International ACM SIGACCESS Conference on Computers \& Accessibility}} (Lisbon, Portugal) \emph{(\bibinfo{series}{ASSETS '15})}. \bibinfo{publisher}{Association for Computing Machinery}, \bibinfo{address}{New York, NY, USA}, \bibinfo{pages}{89–96}.
\newblock
\showISBNx{9781450334006}
\href{https://doi.org/10.1145/2700648.2809851}{doi:\nolinkurl{10.1145/2700648.2809851}}


\bibitem[Fricker(2007)]%
        {fricker_epistemic_2007}
\bibfield{author}{\bibinfo{person}{Miranda Fricker}.} \bibinfo{year}{2007}\natexlab{}.
\newblock \bibinfo{booktitle}{\emph{Epistemic {Injustice}: {Power} and the {Ethics} of {Knowing}}}.
\newblock \bibinfo{publisher}{Clarendon Press}.
\newblock
\showISBNx{978-0-19-823790-7}
\newblock
\shownote{Google-Books-ID: lncSDAAAQBAJ}.


\bibitem[Fuchsberger et~al\mbox{.}(2013)]%
        {Fuchsberger_et_al_13_materiality}
\bibfield{author}{\bibinfo{person}{Verena Fuchsberger}, \bibinfo{person}{Martin Murer}, {and} \bibinfo{person}{Manfred Tscheligi}.} \bibinfo{year}{2013}\natexlab{}.
\newblock \showarticletitle{Materials, materiality, and media}. In \bibinfo{booktitle}{\emph{Proceedings of the SIGCHI Conference on Human Factors in Computing Systems}} (Paris, France) \emph{(\bibinfo{series}{CHI '13})}. \bibinfo{publisher}{Association for Computing Machinery}, \bibinfo{address}{New York, NY, USA}, \bibinfo{pages}{2853–2862}.
\newblock
\showISBNx{9781450318990}
\href{https://doi.org/10.1145/2470654.2481395}{doi:\nolinkurl{10.1145/2470654.2481395}}


\bibitem[Fuchsberger et~al\mbox{.}(2014)]%
        {Fuchsberger_et_al_14}
\bibfield{author}{\bibinfo{person}{Verena Fuchsberger}, \bibinfo{person}{Martin Murer}, {and} \bibinfo{person}{Manfred Tscheligi}.} \bibinfo{year}{2014}\natexlab{}.
\newblock \showarticletitle{Human-Computer Non-Interaction: The Activity of Non-Use}. In \bibinfo{booktitle}{\emph{Proceedings of the 2014 Companion Publication on Designing Interactive Systems}} (Vancouver, BC, Canada) \emph{(\bibinfo{series}{DIS Companion '14})}. \bibinfo{publisher}{Association for Computing Machinery}, \bibinfo{address}{New York, NY, USA}, \bibinfo{pages}{57â€“60}.
\newblock
\showISBNx{9781450329033}
\href{https://doi.org/10.1145/2598784.2602781}{doi:\nolinkurl{10.1145/2598784.2602781}}


\bibitem[Fulton(2019)]%
        {fulton_2019_misophonia}
\bibfield{author}{\bibinfo{person}{April Fulton}.} \bibinfo{year}{2019}\natexlab{}.
\newblock \bibinfo{title}{Misophonia: {When} {Life}'s {Noises} {Drive} {You} {Mad}}.
\newblock
\urldef\tempurl%
\url{https://www.npr.org/sections/health-shots/2019/03/18/702784044/misophonia-when-lifes-noises-drive-you-mad}
\showURL{%
\tempurl}


\bibitem[Garcia et~al\mbox{.}(2020)]%
        {garciaetal2020}
\bibfield{author}{\bibinfo{person}{Patricia Garcia}, \bibinfo{person}{Tonia Sutherland}, \bibinfo{person}{Marika Cifor}, \bibinfo{person}{Anita~Say Chan}, \bibinfo{person}{Lauren Klein}, \bibinfo{person}{Catherine D'Ignazio}, {and} \bibinfo{person}{Niloufar Salehi}.} \bibinfo{year}{2020}\natexlab{}.
\newblock \showarticletitle{No: Critical Refusal as Feminist Data Practice}. In \bibinfo{booktitle}{\emph{Companion Publication of the 2020 Conference on Computer Supported Cooperative Work and Social Computing}} (Virtual Event, USA) \emph{(\bibinfo{series}{CSCW '20 Companion})}. \bibinfo{publisher}{Association for Computing Machinery}, \bibinfo{address}{New York, NY, USA}, \bibinfo{pages}{199–202}.
\newblock
\showISBNx{9781450380591}
\href{https://doi.org/10.1145/3406865.3419014}{doi:\nolinkurl{10.1145/3406865.3419014}}


\bibitem[Gorm and Shklovski(2019)]%
        {gorm2019episodic}
\bibfield{author}{\bibinfo{person}{Nanna Gorm} {and} \bibinfo{person}{Irina Shklovski}.} \bibinfo{year}{2019}\natexlab{}.
\newblock \showarticletitle{Episodic use: Practices of care in self-tracking}.
\newblock \bibinfo{journal}{\emph{New Media \& Society}} \bibinfo{volume}{21}, \bibinfo{number}{11-12} (\bibinfo{year}{2019}), \bibinfo{pages}{2505--2521}.
\newblock


\bibitem[Harper(2007)]%
        {harper2007there}
\bibfield{author}{\bibinfo{person}{Simon Harper}.} \bibinfo{year}{2007}\natexlab{}.
\newblock \showarticletitle{Is there design-for-all?}
\newblock \bibinfo{journal}{\emph{Universal Access in the Information Society}} \bibinfo{volume}{6}, \bibinfo{number}{1} (\bibinfo{year}{2007}), \bibinfo{pages}{111--113}.
\newblock


\bibitem[Hou et~al\mbox{.}(2021)]%
        {hou2021rule}
\bibfield{author}{\bibinfo{person}{Yuanbo Hou}, \bibinfo{person}{Yi Deng}, \bibinfo{person}{Bilei Zhu}, \bibinfo{person}{Zejun Ma}, {and} \bibinfo{person}{Dick Botteldooren}.} \bibinfo{year}{2021}\natexlab{}.
\newblock \showarticletitle{Rule-embedded network for audio-visual voice activity detection in live musical video streams}. In \bibinfo{booktitle}{\emph{ICASSP 2021-2021 IEEE International Conference on Acoustics, Speech and Signal Processing (ICASSP)}}. IEEE, \bibinfo{pages}{4165--4169}.
\newblock


\bibitem[Im et~al\mbox{.}(2021)]%
        {Im_um_2021}
\bibfield{author}{\bibinfo{person}{Jane Im}, \bibinfo{person}{Jill Dimond}, \bibinfo{person}{Melody Berton}, \bibinfo{person}{Una Lee}, \bibinfo{person}{Katherine Mustelier}, \bibinfo{person}{Mark~S. Ackerman}, {and} \bibinfo{person}{Eric Gilbert}.} \bibinfo{year}{2021}\natexlab{}.
\newblock \showarticletitle{Yes: Affirmative Consent as a Theoretical Framework for Understanding and Imagining Social Platforms}. In \bibinfo{booktitle}{\emph{Proceedings of the 2021 CHI Conference on Human Factors in Computing Systems}} (Yokohama, Japan) \emph{(\bibinfo{series}{CHI '21})}. \bibinfo{publisher}{Association for Computing Machinery}, \bibinfo{address}{New York, NY, USA}, Article \bibinfo{articleno}{403}, \bibinfo{numpages}{18}~pages.
\newblock
\showISBNx{9781450380966}
\href{https://doi.org/10.1145/3411764.3445778}{doi:\nolinkurl{10.1145/3411764.3445778}}


\bibitem[Jaswal et~al\mbox{.}(2021)]%
        {jaswal2021misokinesia}
\bibfield{author}{\bibinfo{person}{Sumeet~M Jaswal}, \bibinfo{person}{Andreas~KF De~Bleser}, {and} \bibinfo{person}{Todd~C Handy}.} \bibinfo{year}{2021}\natexlab{}.
\newblock \showarticletitle{Misokinesia is a sensitivity to seeing others fidget that is prevalent in the general population}.
\newblock \bibinfo{journal}{\emph{Scientific reports}} \bibinfo{volume}{11}, \bibinfo{number}{1} (\bibinfo{year}{2021}), \bibinfo{pages}{17204}.
\newblock


\bibitem[Jhaver et~al\mbox{.}(2018)]%
        {Jhaveretal2018}
\bibfield{author}{\bibinfo{person}{Shagun Jhaver}, \bibinfo{person}{Yoni Karpfen}, {and} \bibinfo{person}{Judd Antin}.} \bibinfo{year}{2018}\natexlab{}.
\newblock \showarticletitle{Algorithmic Anxiety and Coping Strategies of Airbnb Hosts}. In \bibinfo{booktitle}{\emph{Proceedings of the 2018 CHI Conference on Human Factors in Computing Systems}} (Montreal QC, Canada) \emph{(\bibinfo{series}{CHI '18})}. \bibinfo{publisher}{Association for Computing Machinery}, \bibinfo{address}{New York, NY, USA}, \bibinfo{pages}{1–12}.
\newblock
\showISBNx{9781450356206}
\href{https://doi.org/10.1145/3173574.3173995}{doi:\nolinkurl{10.1145/3173574.3173995}}


\bibitem[Jhaver et~al\mbox{.}(2023)]%
        {Jhaver_et_al_23}
\bibfield{author}{\bibinfo{person}{Shagun Jhaver}, \bibinfo{person}{Alice~Qian Zhang}, \bibinfo{person}{Quan~Ze Chen}, \bibinfo{person}{Nikhila Natarajan}, \bibinfo{person}{Ruotong Wang}, {and} \bibinfo{person}{Amy~X. Zhang}.} \bibinfo{year}{2023}\natexlab{}.
\newblock \showarticletitle{Personalizing Content Moderation on Social Media: User Perspectives on Moderation Choices, Interface Design, and Labor}.
\newblock \bibinfo{journal}{\emph{Proc. ACM Hum.-Comput. Interact.}} \bibinfo{volume}{7}, \bibinfo{number}{CSCW2}, Article \bibinfo{articleno}{289} (\bibinfo{date}{oct} \bibinfo{year}{2023}), \bibinfo{numpages}{33}~pages.
\newblock
\href{https://doi.org/10.1145/3610080}{doi:\nolinkurl{10.1145/3610080}}


\bibitem[Jhaver and Zhang(2023)]%
        {jhaver2023users}
\bibfield{author}{\bibinfo{person}{Shagun Jhaver} {and} \bibinfo{person}{Amy~X Zhang}.} \bibinfo{year}{2023}\natexlab{}.
\newblock \showarticletitle{Do users want platform moderation or individual control? Examining the role of third-person effects and free speech support in shaping moderation preferences}.
\newblock \bibinfo{journal}{\emph{New Media \& Society}} (\bibinfo{year}{2023}), \bibinfo{pages}{14614448231217993}.
\newblock


\bibitem[Jiang et~al\mbox{.}(2025)]%
        {jiang2025shifting}
\bibfield{author}{\bibinfo{person}{Lucy Jiang}, \bibinfo{person}{Woojin Ko}, \bibinfo{person}{Shirley Yuan}, \bibinfo{person}{Tanisha Shende}, {and} \bibinfo{person}{Shiri Azenkot}.} \bibinfo{year}{2025}\natexlab{}.
\newblock \showarticletitle{Shifting the Focus: Exploring Video Accessibility Strategies and Challenges for People with ADHD}. In \bibinfo{booktitle}{\emph{Proceedings of the 2025 CHI Conference on Human Factors in Computing Systems}} \emph{(\bibinfo{series}{CHI '25})}. \bibinfo{publisher}{Association for Computing Machinery}, \bibinfo{address}{New York, NY, USA}, Article \bibinfo{articleno}{561}, \bibinfo{numpages}{16}~pages.
\newblock
\showISBNx{9798400713941}
\href{https://doi.org/10.1145/3706598.3713637}{doi:\nolinkurl{10.1145/3706598.3713637}}


\bibitem[Johnson-Laird(1989)]%
        {johnson1989mental}
\bibfield{author}{\bibinfo{person}{Philip~N Johnson-Laird}.} \bibinfo{year}{1989}\natexlab{}.
\newblock \showarticletitle{Mental models.}
\newblock  (\bibinfo{year}{1989}).
\newblock


\bibitem[Joy et~al\mbox{.}(2025)]%
        {joyammari25}
\bibfield{author}{\bibinfo{person}{Karen Joy}, \bibinfo{person}{Michelle Liang}, {and} \bibinfo{person}{Tawfiq Ammari}.} \bibinfo{year}{2025}\natexlab{}.
\newblock \showarticletitle{"If it has an exclamation point, I step away from it, I need facts, not excited feelings": Technologically Mediated Parental COVID Uncertainty}.
\newblock \bibinfo{journal}{\emph{Proc. ACM Hum.-Comput. Interact.}} \bibinfo{volume}{9}, \bibinfo{number}{2}, Article \bibinfo{articleno}{CSCW111} (\bibinfo{date}{May} \bibinfo{year}{2025}), \bibinfo{numpages}{38}~pages.
\newblock
\href{https://doi.org/10.1145/3711009}{doi:\nolinkurl{10.1145/3711009}}


\bibitem[Kim and Pardo(2018)]%
        {kim_pardo_2018}
\bibfield{author}{\bibinfo{person}{Bongjun Kim} {and} \bibinfo{person}{Bryan Pardo}.} \bibinfo{year}{2018}\natexlab{}.
\newblock \showarticletitle{A Human-in-the-Loop System for Sound Event Detection and Annotation}.
\newblock \bibinfo{journal}{\emph{ACM Trans. Interact. Intell. Syst.}} \bibinfo{volume}{8}, \bibinfo{number}{2}, Article \bibinfo{articleno}{13} (\bibinfo{date}{jun} \bibinfo{year}{2018}), \bibinfo{numpages}{23}~pages.
\newblock
\showISSN{2160-6455}
\href{https://doi.org/10.1145/3214366}{doi:\nolinkurl{10.1145/3214366}}


\bibitem[Kim et~al\mbox{.}(2024)]%
        {le_et_al_24}
\bibfield{author}{\bibinfo{person}{Bo~Young Kim}, \bibinfo{person}{Qingyan Ma}, {and} \bibinfo{person}{Lisa Diamond}.} \bibinfo{year}{2024}\natexlab{}.
\newblock \showarticletitle{“It’s in My language”: A Case Study on Multilingual mHealth Application for Immigrant Populations With Limited English Proficiency}. In \bibinfo{booktitle}{\emph{Extended Abstracts of the CHI Conference on Human Factors in Computing Systems}} (Honolulu, HI, USA) \emph{(\bibinfo{series}{CHI EA '24})}. \bibinfo{publisher}{Association for Computing Machinery}, \bibinfo{address}{New York, NY, USA}, Article \bibinfo{articleno}{544}, \bibinfo{numpages}{7}~pages.
\newblock
\showISBNx{9798400703317}
\href{https://doi.org/10.1145/3613905.3637125}{doi:\nolinkurl{10.1145/3613905.3637125}}


\bibitem[Kim et~al\mbox{.}(2017)]%
        {kim2017real}
\bibfield{author}{\bibinfo{person}{Juhyun Kim}, \bibinfo{person}{Cheonbok Park}, \bibinfo{person}{Jinwoo Ahn}, \bibinfo{person}{Youlim Ko}, \bibinfo{person}{Junghyun Park}, {and} \bibinfo{person}{John~C Gallagher}.} \bibinfo{year}{2017}\natexlab{}.
\newblock \showarticletitle{Real-time UAV sound detection and analysis system}. In \bibinfo{booktitle}{\emph{2017 IEEE Sensors Applications Symposium (SAS)}}. IEEE, \bibinfo{pages}{1--5}.
\newblock


\bibitem[Li et~al\mbox{.}(2024)]%
        {li2024codesigning}
\bibfield{author}{\bibinfo{person}{Jingjin Li}, \bibinfo{person}{Shaomei Wu}, {and} \bibinfo{person}{Gilly Leshed}.} \bibinfo{year}{2024}\natexlab{}.
\newblock \showarticletitle{Re-envisioning Remote Meetings: Co-designing Inclusive and Empowering Videoconferencing with People Who Stutter}. In \bibinfo{booktitle}{\emph{Proceedings of the 2024 ACM Designing Interactive Systems Conference}} (Copenhagen, Denmark) \emph{(\bibinfo{series}{DIS '24})}. \bibinfo{publisher}{Association for Computing Machinery}, \bibinfo{address}{New York, NY, USA}, \bibinfo{pages}{1926–1941}.
\newblock
\showISBNx{9798400705830}
\href{https://doi.org/10.1145/3643834.3661533}{doi:\nolinkurl{10.1145/3643834.3661533}}


\bibitem[Li et~al\mbox{.}(2023)]%
        {lietalmulti23}
\bibfield{author}{\bibinfo{person}{Yao Li}, \bibinfo{person}{Yubo Kou}, \bibinfo{person}{Renkai Ma}, \bibinfo{person}{Yanlai Wu}, \bibinfo{person}{Guo Freeman}, {and} \bibinfo{person}{Bryan Semaan}.} \bibinfo{year}{2023}\natexlab{}.
\newblock \showarticletitle{Multi-Stakeholder Privacy and Safety on Content Creation Platforms}. In \bibinfo{booktitle}{\emph{Companion Publication of the 2023 ACM Designing Interactive Systems Conference}} (Pittsburgh, PA, USA) \emph{(\bibinfo{series}{DIS '23 Companion})}. \bibinfo{publisher}{Association for Computing Machinery}, \bibinfo{address}{New York, NY, USA}, \bibinfo{pages}{121–123}.
\newblock
\showISBNx{9781450398985}
\href{https://doi.org/10.1145/3563703.3591461}{doi:\nolinkurl{10.1145/3563703.3591461}}


\bibitem[Light and Akama(2012)]%
        {light_akama_2012}
\bibfield{author}{\bibinfo{person}{Ann Light} {and} \bibinfo{person}{Yoko Akama}.} \bibinfo{year}{2012}\natexlab{}.
\newblock \showarticletitle{The human touch: participatory practice and the role of facilitation in designing with communities}. In \bibinfo{booktitle}{\emph{Proceedings of the 12th Participatory Design Conference: Research Papers - Volume 1}} (Roskilde, Denmark) \emph{(\bibinfo{series}{PDC '12})}. \bibinfo{publisher}{Association for Computing Machinery}, \bibinfo{address}{New York, NY, USA}, \bibinfo{pages}{61–70}.
\newblock
\showISBNx{9781450308465}
\href{https://doi.org/10.1145/2347635.2347645}{doi:\nolinkurl{10.1145/2347635.2347645}}


\bibitem[Light et~al\mbox{.}(2018)]%
        {light2018walkthrough}
\bibfield{author}{\bibinfo{person}{Ben Light}, \bibinfo{person}{Jean Burgess}, {and} \bibinfo{person}{Stefanie Duguay}.} \bibinfo{year}{2018}\natexlab{}.
\newblock \showarticletitle{The walkthrough method: An approach to the study of apps}.
\newblock \bibinfo{journal}{\emph{New media \& society}} \bibinfo{volume}{20}, \bibinfo{number}{3} (\bibinfo{year}{2018}), \bibinfo{pages}{881--900}.
\newblock


\bibitem[Lukava et~al\mbox{.}(2022)]%
        {lukava2022two}
\bibfield{author}{\bibinfo{person}{Tamari Lukava}, \bibinfo{person}{Dafne~Zuleima Morgado~Ramirez}, {and} \bibinfo{person}{Giulia Barbareschi}.} \bibinfo{year}{2022}\natexlab{}.
\newblock \showarticletitle{Two sides of the same coin: accessibility practices and neurodivergent users' experience of extended reality}.
\newblock \bibinfo{journal}{\emph{Journal of Enabling Technologies}} \bibinfo{volume}{16}, \bibinfo{number}{2} (\bibinfo{year}{2022}), \bibinfo{pages}{75--90}.
\newblock


\bibitem[Malviya et~al\mbox{.}(2023)]%
        {malviyaetal23}
\bibfield{author}{\bibinfo{person}{Vikas~Kumar Malviya}, \bibinfo{person}{Chee~Wei Leow}, \bibinfo{person}{Ashok Kasthuri}, \bibinfo{person}{Yan~Naing Tun}, \bibinfo{person}{Lwin~Khin Shar}, {and} \bibinfo{person}{Lingxiao Jiang}.} \bibinfo{year}{2023}\natexlab{}.
\newblock \showarticletitle{Right to Know, Right to Refuse: Towards UI Perception-Based Automated Fine-Grained Permission Controls for Android Apps}. In \bibinfo{booktitle}{\emph{Proceedings of the 37th IEEE/ACM International Conference on Automated Software Engineering}} (Rochester, MI, USA) \emph{(\bibinfo{series}{ASE '22})}. \bibinfo{publisher}{Association for Computing Machinery}, \bibinfo{address}{New York, NY, USA}, Article \bibinfo{articleno}{186}, \bibinfo{numpages}{6}~pages.
\newblock
\showISBNx{9781450394758}
\href{https://doi.org/10.1145/3551349.3559556}{doi:\nolinkurl{10.1145/3551349.3559556}}


\bibitem[Michel et~al\mbox{.}(2025)]%
        {mittal_et_al_25}
\bibfield{author}{\bibinfo{person}{Shira Michel}, \bibinfo{person}{Sufi Kaur}, \bibinfo{person}{Sarah~Elizabeth Gillespie}, \bibinfo{person}{Jeffrey Gleason}, \bibinfo{person}{Christo Wilson}, {and} \bibinfo{person}{Avijit Ghosh}.} \bibinfo{year}{2025}\natexlab{}.
\newblock \showarticletitle{“It’s not a representation of me”: Examining Accent Bias and Digital Exclusion in Synthetic AI Voice Services}. In \bibinfo{booktitle}{\emph{Proceedings of the 2025 ACM Conference on Fairness, Accountability, and Transparency}} \emph{(\bibinfo{series}{FAccT '25})}. \bibinfo{publisher}{Association for Computing Machinery}, \bibinfo{address}{New York, NY, USA}, \bibinfo{pages}{228–245}.
\newblock
\showISBNx{9798400714825}
\href{https://doi.org/10.1145/3715275.3732018}{doi:\nolinkurl{10.1145/3715275.3732018}}


\bibitem[Mohamed(2024)]%
        {mohamed2024debilitating}
\bibfield{author}{\bibinfo{person}{Kharnita Mohamed}.} \bibinfo{year}{2024}\natexlab{}.
\newblock \showarticletitle{Debilitating Research: Scholarship of the Obvious and Epistemic Trauma}.
\newblock \bibinfo{journal}{\emph{African Studies}} \bibinfo{volume}{83}, \bibinfo{number}{2-3} (\bibinfo{year}{2024}), \bibinfo{pages}{134--151}.
\newblock


\bibitem[Nevsky et~al\mbox{.}(2025)]%
        {nevskyetal25}
\bibfield{author}{\bibinfo{person}{Alexandre Nevsky}, \bibinfo{person}{Filip Bircanin}, \bibinfo{person}{Elena Simperl}, \bibinfo{person}{Madeline~N Cruice}, {and} \bibinfo{person}{Timothy Neate}.} \bibinfo{year}{2025}\natexlab{}.
\newblock \showarticletitle{To Each Their Own: Exploring Highly Personalised Audiovisual Media Accessibility Interventions with People with Aphasia}. In \bibinfo{booktitle}{\emph{Proceedings of the 2025 ACM Designing Interactive Systems Conference}} \emph{(\bibinfo{series}{DIS '25})}. \bibinfo{publisher}{Association for Computing Machinery}, \bibinfo{address}{New York, NY, USA}, \bibinfo{pages}{1826–1843}.
\newblock
\showISBNx{9798400714856}
\href{https://doi.org/10.1145/3715336.3735771}{doi:\nolinkurl{10.1145/3715336.3735771}}


\bibitem[Pandey et~al\mbox{.}(2023)]%
        {pandey2023nocturnal}
\bibfield{author}{\bibinfo{person}{Chandrasen Pandey}, \bibinfo{person}{Neeraj Baghel}, \bibinfo{person}{Rinki Gupta}, {and} \bibinfo{person}{Malay~Kishore Dutta}.} \bibinfo{year}{2023}\natexlab{}.
\newblock \showarticletitle{Nocturnal sleep sounds classification with artificial neural network for sleep monitoring}.
\newblock \bibinfo{journal}{\emph{Multimedia Tools and Applications}} (\bibinfo{year}{2023}), \bibinfo{pages}{1--17}.
\newblock


\bibitem[Parchoma(2014)]%
        {parchoma_contested_2014}
\bibfield{author}{\bibinfo{person}{Gale Parchoma}.} \bibinfo{year}{2014}\natexlab{}.
\newblock \showarticletitle{The contested ontology of affordances: Implications for researching technological affordances for collaborative knowledge production}.
\newblock \bibinfo{journal}{\emph{Computers in Human Behavior}}  \bibinfo{volume}{37} (\bibinfo{year}{2014}), \bibinfo{pages}{360--368}.
\newblock
\href{https://doi.org/10.1016/j.chb.2012.05.028}{doi:\nolinkurl{10.1016/j.chb.2012.05.028}}


\bibitem[Petronio(2002)]%
        {petronio2002boundaries}
\bibfield{author}{\bibinfo{person}{Sandra Petronio}.} \bibinfo{year}{2002}\natexlab{}.
\newblock \bibinfo{booktitle}{\emph{Boundaries of privacy: Dialectics of disclosure}}.
\newblock \bibinfo{publisher}{Suny Press}.
\newblock


\bibitem[Poerio(2016)]%
        {poerio2016could}
\bibfield{author}{\bibinfo{person}{Giulia Poerio}.} \bibinfo{year}{2016}\natexlab{}.
\newblock \showarticletitle{Could insomnia be relieved with a YouTube video? The relaxation and calm of ASMR}.
\newblock \bibinfo{journal}{\emph{The restless compendium}}  \bibinfo{volume}{119} (\bibinfo{year}{2016}).
\newblock


\bibitem[Potgieter et~al\mbox{.}(2019)]%
        {potgieter2019misophonia}
\bibfield{author}{\bibinfo{person}{Iskra Potgieter}, \bibinfo{person}{Carol MacDonald}, \bibinfo{person}{Lucy Partridge}, \bibinfo{person}{Rilana Cima}, \bibinfo{person}{Jacqueline Sheldrake}, {and} \bibinfo{person}{Derek~J Hoare}.} \bibinfo{year}{2019}\natexlab{}.
\newblock \showarticletitle{Misophonia: A scoping review of research}.
\newblock \bibinfo{journal}{\emph{Journal of clinical psychology}} \bibinfo{volume}{75}, \bibinfo{number}{7} (\bibinfo{year}{2019}), \bibinfo{pages}{1203--1218}.
\newblock


\bibitem[Poulsen et~al\mbox{.}(2024)]%
        {poulsen2024auditory}
\bibfield{author}{\bibinfo{person}{Rebecca Poulsen}, \bibinfo{person}{Z Williams}, \bibinfo{person}{Patrick Dwyer}, \bibinfo{person}{E Pellicano}, \bibinfo{person}{Paul~F Sowman}, {and} \bibinfo{person}{David McAlpine}.} \bibinfo{year}{2024}\natexlab{}.
\newblock \showarticletitle{How auditory processing influences the autistic profile: A review}.
\newblock \bibinfo{journal}{\emph{Autism Research}} \bibinfo{volume}{17}, \bibinfo{number}{12} (\bibinfo{year}{2024}), \bibinfo{pages}{2452--2470}.
\newblock


\bibitem[Race et~al\mbox{.}(2021)]%
        {race_et_al_21}
\bibfield{author}{\bibinfo{person}{Lauren Race}, \bibinfo{person}{Amber James}, \bibinfo{person}{Andrew Hayward}, \bibinfo{person}{Kia El-Amin}, \bibinfo{person}{Maya~Gold Patterson}, {and} \bibinfo{person}{Theresa Mershon}.} \bibinfo{year}{2021}\natexlab{}.
\newblock \showarticletitle{Designing Sensory and Social Tools for Neurodivergent Individuals in Social Media Environments}. In \bibinfo{booktitle}{\emph{Proceedings of the 23rd International ACM SIGACCESS Conference on Computers and Accessibility}} (Virtual Event, USA) \emph{(\bibinfo{series}{ASSETS '21})}. \bibinfo{publisher}{Association for Computing Machinery}, \bibinfo{address}{New York, NY, USA}, Article \bibinfo{articleno}{61}, \bibinfo{numpages}{5}~pages.
\newblock
\showISBNx{9781450383066}
\href{https://doi.org/10.1145/3441852.3476546}{doi:\nolinkurl{10.1145/3441852.3476546}}


\bibitem[Raffaseder and Parker(2008)]%
        {raffaseder2008interrellation}
\bibfield{author}{\bibinfo{person}{Hannes Raffaseder} {and} \bibinfo{person}{Martin Parker}.} \bibinfo{year}{2008}\natexlab{}.
\newblock \showarticletitle{Interrellation: Sound-Transformation and Remixing in Real-Time}. In \bibinfo{booktitle}{\emph{Transdisciplinary Digital Art. Sound, Vision and the New Screen: Digital Art Weeks and Interactive Futures 2006/2007, Zurich, Switzerland and Victoria, BC, Canada. Selected Papers}}. Springer, \bibinfo{pages}{229--237}.
\newblock


\bibitem[Ringland and Wolf(2021)]%
        {ringland_et_al_21}
\bibfield{author}{\bibinfo{person}{Kathryn~E. Ringland} {and} \bibinfo{person}{Christine~T. Wolf}.} \bibinfo{year}{2021}\natexlab{}.
\newblock \showarticletitle{Creating Assistive Technology in Disabled Communities, Five Years on: A Reflection of Neurodivergency and Crafting Accessible Social Spaces}.
\newblock \bibinfo{journal}{\emph{SIGACCESS Access. Comput.}} \bibinfo{number}{131}, Article \bibinfo{articleno}{2} (\bibinfo{date}{dec} \bibinfo{year}{2021}), \bibinfo{numpages}{5}~pages.
\newblock
\showISSN{1558-2337}
\href{https://doi.org/10.1145/3507912.3507914}{doi:\nolinkurl{10.1145/3507912.3507914}}


\bibitem[Robinson et~al\mbox{.}(2020)]%
        {robinsonetal2020}
\bibfield{author}{\bibinfo{person}{Sarah Robinson}, \bibinfo{person}{Nicola J.~Bidwell}, \bibinfo{person}{Laura Maye}, \bibinfo{person}{Nadia Pantidi}, {and} \bibinfo{person}{Conor Linehan}.} \bibinfo{year}{2020}\natexlab{}.
\newblock \showarticletitle{Participation through substituting and refusing}. In \bibinfo{booktitle}{\emph{Proceedings of the 16th Participatory Design Conference 2020 - Participation(s) Otherwise - Volume 2}} (Manizales, Colombia) \emph{(\bibinfo{series}{PDC '20})}. \bibinfo{publisher}{Association for Computing Machinery}, \bibinfo{address}{New York, NY, USA}, \bibinfo{pages}{143–147}.
\newblock
\showISBNx{9781450376068}
\href{https://doi.org/10.1145/3384772.3385148}{doi:\nolinkurl{10.1145/3384772.3385148}}


\bibitem[Sabinson(2025a)]%
        {sabinson2024pictorial}
\bibfield{author}{\bibinfo{person}{Elena Sabinson}.} \bibinfo{year}{2025}\natexlab{a}.
\newblock \showarticletitle{Blowfish Band: A Wearable Inflatable Fidget for Self-Stimulatory (Stim) Behaviors}. In \bibinfo{booktitle}{\emph{Proceedings of the 2025 ACM Designing Interactive Systems Conference}} \emph{(\bibinfo{series}{DIS '25})}. \bibinfo{publisher}{Association for Computing Machinery}, \bibinfo{address}{New York, NY, USA}, \bibinfo{pages}{1–0}.
\newblock
\showISBNx{9798400714856}
\href{https://doi.org/10.1145/3715336.3735442}{doi:\nolinkurl{10.1145/3715336.3735442}}


\bibitem[Sabinson(2025b)]%
        {sabinson2025blowfish}
\bibfield{author}{\bibinfo{person}{Elena Sabinson}.} \bibinfo{year}{2025}\natexlab{b}.
\newblock \showarticletitle{Blowfish Band: A Wearable Inflatable Fidget for Self-Stimulatory (Stim) Behaviors}. In \bibinfo{booktitle}{\emph{Proceedings of the 2025 ACM Designing Interactive Systems Conference}} \emph{(\bibinfo{series}{DIS '25})}. \bibinfo{publisher}{Association for Computing Machinery}, \bibinfo{address}{New York, NY, USA}, \bibinfo{pages}{1–0}.
\newblock
\showISBNx{9798400714856}
\href{https://doi.org/10.1145/3715336.3735442}{doi:\nolinkurl{10.1145/3715336.3735442}}


\bibitem[Salovaara et~al\mbox{.}(2011)]%
        {salovaara2011everyday}
\bibfield{author}{\bibinfo{person}{Antti Salovaara}, \bibinfo{person}{Sacha Helfenstein}, {and} \bibinfo{person}{Antti Oulasvirta}.} \bibinfo{year}{2011}\natexlab{}.
\newblock \showarticletitle{Everyday appropriations of information technology: A study of creative uses of digital cameras}.
\newblock \bibinfo{journal}{\emph{Journal of the American Society for Information Science and Technology}} \bibinfo{volume}{62}, \bibinfo{number}{12} (\bibinfo{year}{2011}), \bibinfo{pages}{2347--2363}.
\newblock


\bibitem[Sampson et~al\mbox{.}(2023)]%
        {sampsondenae23}
\bibfield{author}{\bibinfo{person}{Princess Sampson}, \bibinfo{person}{Ro Encarnacion}, {and} \bibinfo{person}{Dana\"{e} Metaxa}.} \bibinfo{year}{2023}\natexlab{}.
\newblock \showarticletitle{Representation, Self-Determination, and Refusal: Queer People’s Experiences with Targeted Advertising}. In \bibinfo{booktitle}{\emph{Proceedings of the 2023 ACM Conference on Fairness, Accountability, and Transparency}} (Chicago, IL, USA) \emph{(\bibinfo{series}{FAccT '23})}. \bibinfo{publisher}{Association for Computing Machinery}, \bibinfo{address}{New York, NY, USA}, \bibinfo{pages}{1711–1722}.
\newblock
\showISBNx{9798400701924}
\href{https://doi.org/10.1145/3593013.3594110}{doi:\nolinkurl{10.1145/3593013.3594110}}


\bibitem[Schr{\"o}der et~al\mbox{.}(2013)]%
        {schroder2013misophonia}
\bibfield{author}{\bibinfo{person}{Arjan Schr{\"o}der}, \bibinfo{person}{Nienke Vulink}, {and} \bibinfo{person}{Damiaan Denys}.} \bibinfo{year}{2013}\natexlab{}.
\newblock \showarticletitle{Misophonia: diagnostic criteria for a new psychiatric disorder}.
\newblock \bibinfo{journal}{\emph{PloS one}} \bibinfo{volume}{8}, \bibinfo{number}{1} (\bibinfo{year}{2013}), \bibinfo{pages}{e54706}.
\newblock


\bibitem[Scott et~al\mbox{.}(2023)]%
        {Scott2023}
\bibfield{author}{\bibinfo{person}{Carol~F Scott}, \bibinfo{person}{Gabriela Marcu}, \bibinfo{person}{Riana~Elyse Anderson}, \bibinfo{person}{Mark~W Newman}, {and} \bibinfo{person}{Sarita Schoenebeck}.} \bibinfo{year}{2023}\natexlab{}.
\newblock \showarticletitle{Trauma-Informed Social Media: Towards Solutions for Reducing and Healing Online Harm}. In \bibinfo{booktitle}{\emph{Proceedings of the 2023 CHI Conference on Human Factors in Computing Systems}} (Hamburg, Germany) \emph{(\bibinfo{series}{CHI '23})}. \bibinfo{publisher}{Association for Computing Machinery}, \bibinfo{address}{New York, NY, USA}, Article \bibinfo{articleno}{341}, \bibinfo{numpages}{20}~pages.
\newblock
\showISBNx{9781450394215}
\href{https://doi.org/10.1145/3544548.3581512}{doi:\nolinkurl{10.1145/3544548.3581512}}


\bibitem[Sheehan and Le~Dantec(2023)]%
        {sheehan_23}
\bibfield{author}{\bibinfo{person}{Alyssa Sheehan} {and} \bibinfo{person}{Christopher~A. Le~Dantec}.} \bibinfo{year}{2023}\natexlab{}.
\newblock \showarticletitle{Making Meaning from the Digitalization of Blue-Collar Work}.
\newblock \bibinfo{journal}{\emph{Proc. ACM Hum.-Comput. Interact.}} \bibinfo{volume}{7}, \bibinfo{number}{CSCW2}, Article \bibinfo{articleno}{345} (\bibinfo{date}{Oct.} \bibinfo{year}{2023}), \bibinfo{numpages}{21}~pages.
\newblock
\href{https://doi.org/10.1145/3610194}{doi:\nolinkurl{10.1145/3610194}}


\bibitem[Silverstone et~al\mbox{.}(1994)]%
        {silverstone1992information}
\bibfield{author}{\bibinfo{person}{Roger Silverstone}, \bibinfo{person}{Eric Hirsch}, {and} \bibinfo{person}{David Morley}.} \bibinfo{year}{1994}\natexlab{}.
\newblock \showarticletitle{Information and communication technologies and the moral economy of the household}.
\newblock In \bibinfo{booktitle}{\emph{Consuming Technologies: Media and Information in Domestic Spaces}}, \bibfield{editor}{\bibinfo{person}{Roger Silverstone} {and} \bibinfo{person}{Eric Hirsch}} (Eds.). \bibinfo{publisher}{Routledge}, \bibinfo{pages}{13--28}.
\newblock


\bibitem[S{\o}rensen(2006)]%
        {sorensen2006domestication}
\bibfield{author}{\bibinfo{person}{Knut~H S{\o}rensen}.} \bibinfo{year}{2006}\natexlab{}.
\newblock \showarticletitle{Domestication: the enactment of technology}.
\newblock \bibinfo{journal}{\emph{Domestication of media and technology}}  \bibinfo{volume}{46} (\bibinfo{year}{2006}).
\newblock


\bibitem[Spiel et~al\mbox{.}(2019)]%
        {spiel2019agency}
\bibfield{author}{\bibinfo{person}{Katta Spiel}, \bibinfo{person}{Christopher Frauenberger}, \bibinfo{person}{Os Keyes}, {and} \bibinfo{person}{Geraldine Fitzpatrick}.} \bibinfo{year}{2019}\natexlab{}.
\newblock \showarticletitle{Agency of Autistic Children in Technology Research—A Critical Literature Review}.
\newblock \bibinfo{journal}{\emph{ACM Trans. Comput.-Hum. Interact.}} \bibinfo{volume}{26}, \bibinfo{number}{6}, Article \bibinfo{articleno}{38} (\bibinfo{date}{Nov.} \bibinfo{year}{2019}), \bibinfo{numpages}{40}~pages.
\newblock
\showISSN{1073-0516}
\href{https://doi.org/10.1145/3344919}{doi:\nolinkurl{10.1145/3344919}}


\bibitem[Spiel et~al\mbox{.}(2020)]%
        {spiel2020nothing}
\bibfield{author}{\bibinfo{person}{Katta Spiel}, \bibinfo{person}{Kathrin Gerling}, \bibinfo{person}{Cynthia~L. Bennett}, \bibinfo{person}{Emeline Brul\'{e}}, \bibinfo{person}{Rua~M. Williams}, \bibinfo{person}{Jennifer Rode}, {and} \bibinfo{person}{Jennifer Mankoff}.} \bibinfo{year}{2020}\natexlab{}.
\newblock \showarticletitle{Nothing About Us Without Us: Investigating the Role of Critical Disability Studies in HCI}. In \bibinfo{booktitle}{\emph{Extended Abstracts of the 2020 CHI Conference on Human Factors in Computing Systems}} (Honolulu, HI, USA) \emph{(\bibinfo{series}{CHI EA '20})}. \bibinfo{publisher}{Association for Computing Machinery}, \bibinfo{address}{New York, NY, USA}, \bibinfo{pages}{1–8}.
\newblock
\showISBNx{9781450368193}
\href{https://doi.org/10.1145/3334480.3375150}{doi:\nolinkurl{10.1145/3334480.3375150}}


\bibitem[Spiel et~al\mbox{.}(2022)]%
        {spiel_et_al_22}
\bibfield{author}{\bibinfo{person}{Katta Spiel}, \bibinfo{person}{Eva Hornecker}, \bibinfo{person}{Rua~Mae Williams}, {and} \bibinfo{person}{Judith Good}.} \bibinfo{year}{2022}\natexlab{}.
\newblock \showarticletitle{ADHD and Technology Research – Investigated by Neurodivergent Readers}. In \bibinfo{booktitle}{\emph{Proceedings of the 2022 CHI Conference on Human Factors in Computing Systems}} (New Orleans, LA, USA) \emph{(\bibinfo{series}{CHI '22})}. \bibinfo{publisher}{Association for Computing Machinery}, \bibinfo{address}{New York, NY, USA}, Article \bibinfo{articleno}{547}, \bibinfo{numpages}{21}~pages.
\newblock
\showISBNx{9781450391573}
\href{https://doi.org/10.1145/3491102.3517592}{doi:\nolinkurl{10.1145/3491102.3517592}}


\bibitem[Stein et~al\mbox{.}(2021)]%
        {stein2021mental}
\bibfield{author}{\bibinfo{person}{Dan~J Stein}, \bibinfo{person}{Andrea~C Palk}, {and} \bibinfo{person}{Kenneth~S Kendler}.} \bibinfo{year}{2021}\natexlab{}.
\newblock \showarticletitle{What is a mental disorder? An exemplar-focused approach}.
\newblock \bibinfo{journal}{\emph{Psychological Medicine}} \bibinfo{volume}{51}, \bibinfo{number}{6} (\bibinfo{year}{2021}), \bibinfo{pages}{894--901}.
\newblock


\bibitem[Stephens et~al\mbox{.}(2023)]%
        {stephens2023accessibility}
\bibfield{author}{\bibinfo{person}{Lindsay Stephens}, \bibinfo{person}{Hilda Smith}, \bibinfo{person}{Iris Epstein}, \bibinfo{person}{Melanie Baljko}, \bibinfo{person}{Ian Mcintosh}, \bibinfo{person}{Nastaran Dadashi}, {and} \bibinfo{person}{Devika~Narayani Prakash}.} \bibinfo{year}{2023}\natexlab{}.
\newblock \showarticletitle{Accessibility and participatory design: time, power, and facilitation}.
\newblock \bibinfo{journal}{\emph{CoDesign}} \bibinfo{volume}{19}, \bibinfo{number}{4} (\bibinfo{year}{2023}), \bibinfo{pages}{287--303}.
\newblock


\bibitem[Stinnett(2018)]%
        {stinnett2018trauma}
\bibfield{author}{\bibinfo{person}{Gina Stinnett}.} \bibinfo{year}{2018}\natexlab{}.
\newblock \bibinfo{booktitle}{\emph{Trauma and the Credibility Economy: An Analysis of Epistemic Violence and Its Traumatic Functions}}.
\newblock \bibinfo{publisher}{Illinois State University}.
\newblock


\bibitem[Sum et~al\mbox{.}(2022)]%
        {sum_et_al_22}
\bibfield{author}{\bibinfo{person}{Cella~M Sum}, \bibinfo{person}{Rahaf Alharbi}, \bibinfo{person}{Franchesca Spektor}, \bibinfo{person}{Cynthia~L Bennett}, \bibinfo{person}{Christina~N Harrington}, \bibinfo{person}{Katta Spiel}, {and} \bibinfo{person}{Rua~Mae Williams}.} \bibinfo{year}{2022}\natexlab{}.
\newblock \showarticletitle{Dreaming Disability Justice in HCI}. In \bibinfo{booktitle}{\emph{Extended Abstracts of the 2022 CHI Conference on Human Factors in Computing Systems}} (New Orleans, LA, USA) \emph{(\bibinfo{series}{CHI EA '22})}. \bibinfo{publisher}{Association for Computing Machinery}, \bibinfo{address}{New York, NY, USA}, Article \bibinfo{articleno}{114}, \bibinfo{numpages}{5}~pages.
\newblock
\showISBNx{9781450391566}
\href{https://doi.org/10.1145/3491101.3503731}{doi:\nolinkurl{10.1145/3491101.3503731}}


\bibitem[Swedo et~al\mbox{.}(2022)]%
        {swedo2022consensus}
\bibfield{author}{\bibinfo{person}{Susan~E Swedo}, \bibinfo{person}{David~M Baguley}, \bibinfo{person}{Damiaan Denys}, \bibinfo{person}{Laura~J Dixon}, \bibinfo{person}{Mercede Erfanian}, \bibinfo{person}{Alessandra Fioretti}, \bibinfo{person}{Pawel~J Jastreboff}, \bibinfo{person}{Sukhbinder Kumar}, \bibinfo{person}{M~Zachary Rosenthal}, \bibinfo{person}{Romke Rouw}, {et~al\mbox{.}}} \bibinfo{year}{2022}\natexlab{}.
\newblock \showarticletitle{Consensus definition of misophonia: a delphi study}.
\newblock \bibinfo{journal}{\emph{Frontiers in Neuroscience}}  \bibinfo{volume}{16} (\bibinfo{year}{2022}), \bibinfo{pages}{841816}.
\newblock


\bibitem[Taylor(2017)]%
        {taylor2017misophonia}
\bibfield{author}{\bibinfo{person}{Steven Taylor}.} \bibinfo{year}{2017}\natexlab{}.
\newblock \showarticletitle{Misophonia: A new mental disorder?}
\newblock \bibinfo{journal}{\emph{Medical Hypotheses}}  \bibinfo{volume}{103} (\bibinfo{year}{2017}), \bibinfo{pages}{109--117}.
\newblock


\bibitem[Wall and Gannon-Leary(1999)]%
        {wall1999sentence}
\bibfield{author}{\bibinfo{person}{Celia~J Wall} {and} \bibinfo{person}{Pat Gannon-Leary}.} \bibinfo{year}{1999}\natexlab{}.
\newblock \showarticletitle{A sentence made by men: Muted group theory revisited}.
\newblock \bibinfo{journal}{\emph{European Journal of Women's Studies}} \bibinfo{volume}{6}, \bibinfo{number}{1} (\bibinfo{year}{1999}), \bibinfo{pages}{21--29}.
\newblock


\bibitem[Webb(2022)]%
        {webb2022beta}
\bibfield{author}{\bibinfo{person}{Jadon Webb}.} \bibinfo{year}{2022}\natexlab{}.
\newblock \showarticletitle{$\beta$-Blockers for the Treatment of Misophonia and Misokinesia}.
\newblock \bibinfo{journal}{\emph{Clinical Neuropharmacology}} \bibinfo{volume}{45}, \bibinfo{number}{1} (\bibinfo{year}{2022}), \bibinfo{pages}{13--14}.
\newblock


\bibitem[Wong-Villacres et~al\mbox{.}(2019)]%
        {wong-villacres_et_al_19}
\bibfield{author}{\bibinfo{person}{Marisol Wong-Villacres}, \bibinfo{person}{Neha Kumar}, {and} \bibinfo{person}{Betsy DiSalvo}.} \bibinfo{year}{2019}\natexlab{}.
\newblock \showarticletitle{The Parenting Actor-Network of Latino Immigrants in the United States}. In \bibinfo{booktitle}{\emph{Proceedings of the 2019 CHI Conference on Human Factors in Computing Systems}} (Glasgow, Scotland Uk) \emph{(\bibinfo{series}{CHI '19})}. \bibinfo{publisher}{Association for Computing Machinery}, \bibinfo{address}{New York, NY, USA}, \bibinfo{pages}{1–12}.
\newblock
\showISBNx{9781450359702}
\href{https://doi.org/10.1145/3290605.3300914}{doi:\nolinkurl{10.1145/3290605.3300914}}


\bibitem[Yuan et~al\mbox{.}(2025)]%
        {yuan_et_al_24}
\bibfield{author}{\bibinfo{person}{Shuai Yuan}, \bibinfo{person}{Simon Coghlan}, \bibinfo{person}{Reeva Lederman}, {and} \bibinfo{person}{Jenny Waycott}.} \bibinfo{year}{2025}\natexlab{}.
\newblock \showarticletitle{Language Barriers in Long-term Aged Care Homes: Design Considerations for Translation Technology}. In \bibinfo{booktitle}{\emph{Proceedings of the 36th Australasian Conference on Human-Computer Interaction}} \emph{(\bibinfo{series}{OzCHI '24})}. \bibinfo{publisher}{Association for Computing Machinery}, \bibinfo{address}{New York, NY, USA}, \bibinfo{pages}{134–146}.
\newblock
\showISBNx{9798400715099}
\href{https://doi.org/10.1145/3726986.3727901}{doi:\nolinkurl{10.1145/3726986.3727901}}


\bibitem[Zhao(2023)]%
        {zhao_et_al_23}
\bibfield{author}{\bibinfo{person}{Shichao Zhao}.} \bibinfo{year}{2023}\natexlab{}.
\newblock \showarticletitle{Involving British-Chinese Immigrants in Participatory Action Research: Lessons Learnt from the Field}. In \bibinfo{booktitle}{\emph{Proceedings of the 2023 ACM Designing Interactive Systems Conference}} (Pittsburgh, PA, USA) \emph{(\bibinfo{series}{DIS '23})}. \bibinfo{publisher}{Association for Computing Machinery}, \bibinfo{address}{New York, NY, USA}, \bibinfo{pages}{45–60}.
\newblock
\showISBNx{9781450398930}
\href{https://doi.org/10.1145/3563657.3596107}{doi:\nolinkurl{10.1145/3563657.3596107}}


\bibitem[Zolyomi and Snyder(2021)]%
        {zolyomi2021social}
\bibfield{author}{\bibinfo{person}{Annuska Zolyomi} {and} \bibinfo{person}{Jaime Snyder}.} \bibinfo{year}{2021}\natexlab{}.
\newblock \showarticletitle{Social-Emotional-Sensory Design Map for Affective Computing Informed by Neurodivergent Experiences}.
\newblock \bibinfo{journal}{\emph{Proc. ACM Hum.-Comput. Interact.}} \bibinfo{volume}{5}, \bibinfo{number}{CSCW1}, Article \bibinfo{articleno}{77} (\bibinfo{date}{April} \bibinfo{year}{2021}), \bibinfo{numpages}{37}~pages.
\newblock
\href{https://doi.org/10.1145/3449151}{doi:\nolinkurl{10.1145/3449151}}


\bibitem[Zong and Matias(2024)]%
        {Zong_Matias_24}
\bibfield{author}{\bibinfo{person}{Jonathan Zong} {and} \bibinfo{person}{J.~Nathan Matias}.} \bibinfo{year}{2024}\natexlab{}.
\newblock \showarticletitle{Data Refusal from Below: A Framework for Understanding, Evaluating, and Envisioning Refusal as Design}.
\newblock \bibinfo{journal}{\emph{ACM J. Responsib. Comput.}} \bibinfo{volume}{1}, \bibinfo{number}{1}, Article \bibinfo{articleno}{10} (\bibinfo{date}{March} \bibinfo{year}{2024}), \bibinfo{numpages}{23}~pages.
\newblock
\href{https://doi.org/10.1145/3630107}{doi:\nolinkurl{10.1145/3630107}}


\end{thebibliography}


\end{document}